\documentclass[letterpaper,twocolumn,prl,aps,superscriptaddress,amsmath,amssymb,floatfix]{revtex4-1}
\usepackage{mathptmx}
\usepackage[latin9]{inputenc}
\usepackage{color}
\usepackage{textcomp}
\usepackage{amsmath}
\usepackage{amssymb}
\usepackage{graphicx}
\usepackage{esint}
\usepackage[unicode=true,
 bookmarks=true,bookmarksnumbered=false,bookmarksopen=false,
 breaklinks=false,pdfborder={0 0 0},pdfborderstyle={},backref=false,colorlinks=true]
 {hyperref}
\hypersetup{
 linkcolor=magenta,urlcolor=blue,citecolor=blue,pdfstartview={FitH},hyperfootnotes=false}

\makeatletter

\pdfpageheight\paperheight
\pdfpagewidth\paperwidth

\usepackage{epsfig}\usepackage{amsfonts}

\usepackage{textcomp}
\usepackage{epstopdf}

\usepackage{amsfonts}

\pdfpageheight\paperheight
\pdfpagewidth\paperwidth

\@ifundefined{textcolor}{}{%
 \definecolor{BLACK}{gray}{0}
 \definecolor{WHITE}{gray}{1}
 \definecolor{RED}{rgb}{1,0,0}
 \definecolor{GREEN}{rgb}{0,1,0}
 \definecolor{BLUE}{rgb}{0,0,1}
 \definecolor{CYAN}{cmyk}{1,0,0,0}
 \definecolor{MAGENTA}{cmyk}{0,1,0,0}
 \definecolor{YELLOW}{cmyk}{0,0,1,0}
}

\usepackage{xcolor}\usepackage{soul}
\makeatother

\begin{document}
\title{Enhanced optomechanical entanglement and cooling via dissipation engineering}
\author{Yan-Lei Zhang}
\affiliation{CAS Key Laboratory of Quantum Information, University of Science and
Technology of China, Hefei, Anhui 230026, China }
\affiliation{CAS Center For Excellence in Quantum Information and Quantum Physics, University of
Science and Technology of China, Hefei, Anhui 230026, China}
\author{Chuan-Sheng Yang}
\affiliation{CAS Key Laboratory of Quantum Information, University of Science and
Technology of China, Hefei, Anhui 230026, China }
\affiliation{CAS Center For Excellence in Quantum Information and Quantum Physics, University of 
Science and Technology of China, Hefei, Anhui 230026, China}
\author{Zhen Shen}
\affiliation{CAS Key Laboratory of Quantum Information, University of Science and
Technology of China, Hefei, Anhui 230026, China }
\affiliation{CAS Center For Excellence in Quantum Information and Quantum Physics, University of
Science and Technology of China, Hefei, Anhui 230026, China}
\author{Chun-Hua Dong}
\affiliation{CAS Key Laboratory of Quantum Information, University of Science and
Technology of China, Hefei, Anhui 230026, China }
\affiliation{CAS Center For Excellence in Quantum Information and Quantum Physics, University of
Science and Technology of China, Hefei, Anhui 230026, China}
\author{Guang-Can Guo}
\affiliation{CAS Key Laboratory of Quantum Information, University of Science and
Technology of China, Hefei, Anhui 230026, China }
\affiliation{CAS Center For Excellence in Quantum Information and Quantum Physics, University of
Science and Technology of China, Hefei, Anhui 230026, China}
\author{Chang-Ling Zou}
\email{clzou321@ustc.edu.cn}

\affiliation{CAS Key Laboratory of Quantum Information, University of Science and
Technology of China, Hefei, Anhui 230026, China }
\affiliation{CAS Center For Excellence in Quantum Information and Quantum Physics, University of
Science and Technology of China, Hefei, Anhui 230026, China}
\author{Xu-Bo Zou}
\email{xbz@ustc.edu.cn}

\affiliation{CAS Key Laboratory of Quantum Information, University of Science and
Technology of China, Hefei, Anhui 230026, China }
\affiliation{CAS Center For Excellence in Quantum Information and Quantum Physics, University of
Science and Technology of China, Hefei, Anhui 230026, China}
\date{\today}
\begin{abstract}
We propose an optomechanical dissipation engineering scheme by introducing
an ancillary mechanical mode with a large decay rate to control the
density of states of the optical mode. The effective linewidth of
the optical mode can be reduced or broadened, manifesting the dissipation
engineering. To prove the ability of our scheme in improving the performances
of the optomechanical system, we studied optomechanical entanglement
and phonon cooling. It is demonstrated that the optomechanical entanglement
overwhelmed by thermal phonon excitations could be restored via dissipation
engineering. For the phonon cooling, an order of magnitude improvement
could be achieved. Our scheme can be generalized to other systems
with multiple bosonic modes, which is experimentally feasible with
advances in materials and nanofabrication, including optical Fabry-Perot
cavities, superconducting circuits, and nanobeam photonic crystals.
\end{abstract}
\maketitle

\section{Introduction}

In recent years, optomechanical systems \citep{aspelmeyer2012quantum,kippenberg2008cavity,van2010optomechanical,aspelmeyer2014cavity}
have attracted considerable attention in classical and quantum information
processing. Dramatic theoretical and experimental progress has been
achieved in the optomechanically induced transparency \citep{weis2010optomechanically,safavi2011electromagnetically},
sensing \citep{aspelmeyer2014cavity}, frequency combs \citep{butsch2014cw},
and tunable optical filters \citep{deotare2012all}. For potential
applications in the quantum regime \citep{stannigel2012optomechanical,aspelmeyer2012quantum,vanner2011pulsed,anetsberger2010measuring},
ground-state cooling of the mechanical resonator \citep{teufel2011sideband,meenehan2015pulsed},
optomechanical entanglement and squeezing \citep{vitali2007optomechanical,yang2019optomechanically,lu2015steady},
and quantum sensors \citep{branford2018fundamental,arcizet2006high}
have been studied. Most researches have focused on canonical systems
with only one optical and mechanical degree of freedom. For multimode
optomechanical systems \citep{lee2015multimode,fan2015cascaded,duggan2019optomechanically,wei2019controllable},
more degrees of freedom can be controlled, and new phenomena are observed
and new functionalities are realized, such as optomechanical induced
nonreciprocity \citep{shen2016experimental}, circulators \citep{ruesink2018optical,shen2018reconfigurable},
and coherent frequency convertor \citep{dong2012optomechanical,hill2012coherent}.
However, the interplay between different optomechanical interactions
and different mechanical or optical modes are omitted in previous
studies.

On the other hand, dissipation engineering \citep{verstraete2009quantum,poyatos1996quantum,stannigel2012driven,sarlette2011stabilization}
has been developed in quantum optics. By introducing dissipative modes
into the quantum system, the density of states of the system can be
efficiently controlled and thus suppresses unwanted physical processes
or induces desired interactions. Recently, dissipation engineering
has been widely used to generate steady entanglement in superconducting
qubits \citep{leghtas2013stabilizing,ma2019stabilizing}, trapped
ions \citep{barreiro2011open,schindler2013quantum}, superconducting
resonators \citep{li2012engineering}, and Rydberg atoms \citep{li2018engineering}.
Therefore, it is expected to introduce this beneficial idea to the
optomechanical system. In practical, most optomechanical systems naturally
support the multimodes \citep{aspelmeyer2014cavity,massel2012multimode,deng2016optimizing,ockeloen2019sideband},
and these modes can be excited by polychromatic laser drive \citep{zhang2017optomechanical}.
With the advantages of multiple degrees in multimode optomechanical
systems \citep{nielsen2017multimode}, dissipation engineering \citep{meystre2013short}
holds huge potential in all-optical information processing \citep{lee2018quantum,leghtas2013stabilizing,chen2017dissipative},
where two or more mechanical or optical modes lead to a wealth of
different possible schemes.

In this paper, we theoretically analyze a multimode optomechanical
system, where Brillouin phonon mode and breath mechanical mode are
coupled with a common optical mode. The Brillouin phonon mode serves
as an ancillary, which is used to engineer the optical density of
state and thus realize the dissipation engineering. The effect of
dissipation engineering is studied in different aspects, including
optomechanical induced transparency (OMIT) and optomechanical induced
amplification (OMIA), entanglement between the optical mode and breath
mechanical mode, and cooling of the breath mechanical mode. The enhancement
of the entanglement or cooling is obtained for certain dissipation
engineering strength, and the numerical results reveal the potential
of our scheme. On the other hand, the extra noises introduced by the
ancillary mode and strong modification of the optical mode could also
suppress the performance of an optomechanical system, suggesting an
optimal dissipation engineering that balances the positive and negative
effects. Our scheme could be generalized to other nonlinear interactions
in multiple mode systems, such as atom ensemble and Fabry-Perot (FP)
cavity system \citep{hu2019cavity} and superconducting circuits \citep{devoret2013superconducting},
and finds applications in quantum devices \citep{wallquist2009hybrid}.

\section{The system}

\begin{figure}
\center \includegraphics[width=1\columnwidth]{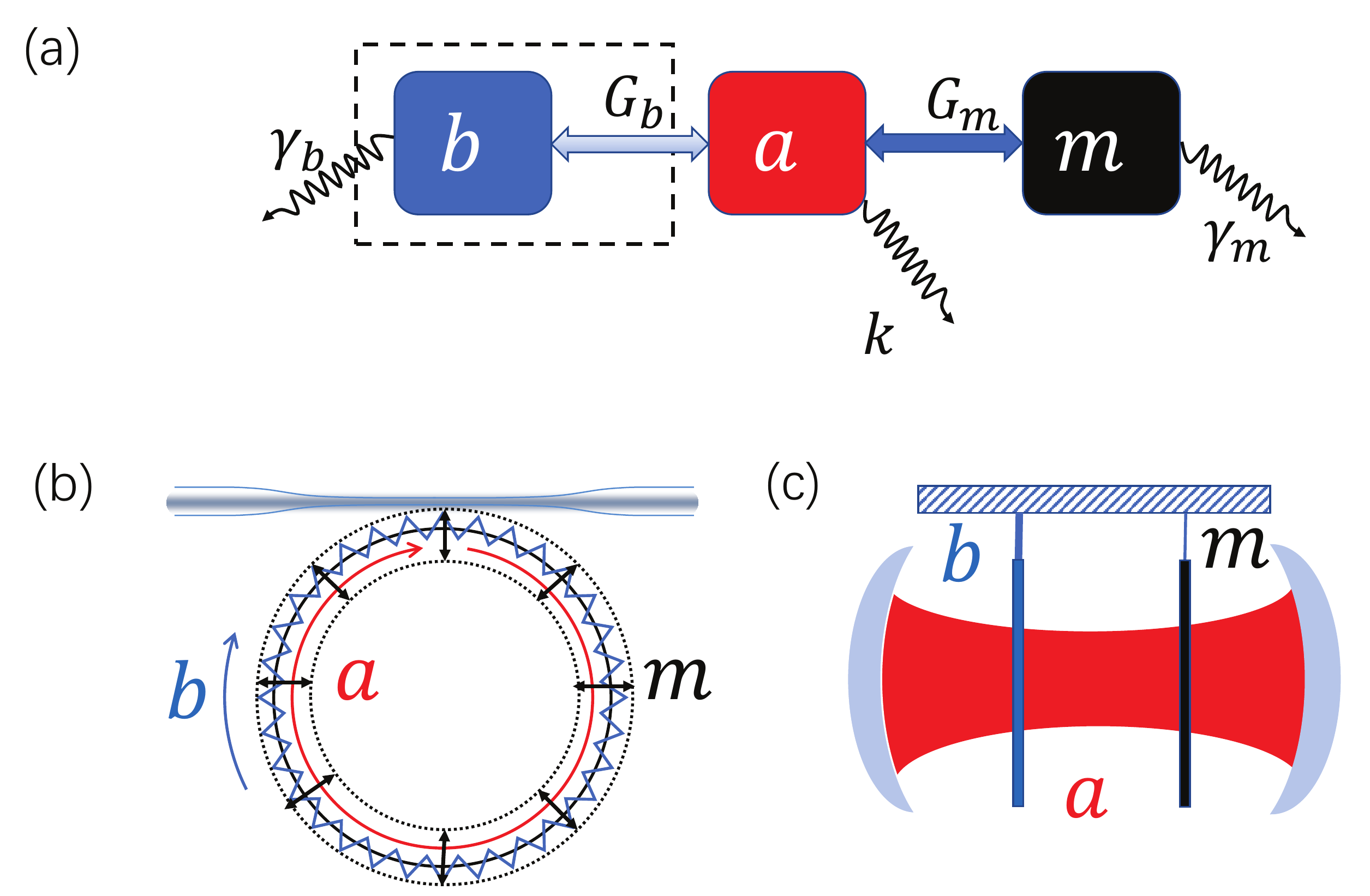} \caption{(Color online) (a) Schematic of the multimode optomechanical system
for dissipation engineering. Potential experimental systems: (b) the
optical microresonator, where both the Brillouin mechanical mode $b$
and the breath mechanical mode $m$ are coupling to the common optical
mode $a$; (c) the Fabry-Perot cavity with two mechanical membranes
inside, the optical mode $a$ can simultaneously couple with mechanical
modes $b$ and $m$ in different membranes.}

\label{Fig1}
\end{figure}
We consider a multimode optomechanical system \citep{shen2018reconfigurable,wang2013reservoir,tian2013robust},
and its schematic is shown in Fig.~\ref{Fig1}(a), where two mechanical
modes $\left(b,~m\right)$ are coupled with a common optical mode
$a$. The interaction Hamiltonian can be described by ($\hbar=1$):
\begin{align}
H_{\mathrm{int}} & =G_{b}b\left(a^{\dagger}+a\right)+G_{m}a\left(m^{\dagger}+m\right)+\mathrm{H.c.},
\end{align}
where $G_{b\left(m\right)}$ is the pump laser stimulated effective
coupling between the modes. Here, we focus on the modes $a,~m$, and
the mode $b$ is treated as an ancillary mode, which is encircled
with a dashed line. By the interaction $G_{b}$, we can modulate the
mode $a$, which further affects the effective coupling between the
modes $a$ and $m$. The dissipation engineering scheme can be realized
in many physical systems, in which two potential systems are the optical
microresonator \citep{shen2018reconfigurable} and FP cavity \citep{hartmann2008steady},
as shown in Figs.~\ref{Fig1}(b) and (c), respectively. In the FP
cavity, the scheme can be easily realized with two driving lasers,
and two mechanical modes $b,~m$ are coupled with the optical mode
$a$ by the red or blue detuned pump lasers. The scheme can be also
generalized to other bosonic systems, including magnon \citep{zhang2016optomagnonic},
atomic spin excitation \citep{pezze2018quantum}, and photonic crystals
\citep{viasnoff2005compact}.

In the following, we focus on an experimentally optomechanical system
based on the on-chip optical microresonator, where the coherent couplings
between the optical mode and the Brillouin phonon mode ($\sim10\mathrm{GHz}$)
\citep{dong2015brillouin}, breath mechanical mode ($\sim100\mathrm{MHz}$)
\citep{shen2016experimental} can be stimulated simultaneously by
external laser pumps. The full system can be described by the following
Hamiltonian
\begin{align}
H_{\mathrm{sys}}= & H_{\mathrm{0}}+H_{\mathrm{SBS}}+H_{\mathrm{BM}}+H_{\mathrm{D}},
\end{align}
where
\begin{align}
H_{\mathrm{0}} & =\omega_{j}a_{j}^{\dagger}a_{j}+\omega_{k}a_{k}^{\dagger}a_{k}+\omega_{b}b_{j-k}^{\dagger}b_{j-k}+\omega_{m}m^{\dagger}m,\\
H_{\mathrm{SBS}} & =g_{b}\left(a_{j}^{\dagger}a_{k}b_{j-k}+a_{j}a_{k}^{\dagger}b_{j-k}^{\dagger}\right),\\
H_{\mathrm{BM}} & =g_{m,j}a_{j}^{\dagger}a_{j}\left(m+m^{\dagger}\right)+g_{m,k}a_{k}^{\dagger}a_{k}\left(m+m^{\dagger}\right),\\
H_{\mathrm{D}} & =i\sqrt{\kappa_{j,ex}}\epsilon_{d,j}\left(a_{j}^{\dagger}e^{-i\omega_{d,j}t}-a_{j}e^{i\omega_{d,j}t}\right)\nonumber \\
 & +i\sqrt{\kappa_{k,ex}}\epsilon_{d,k}\left(a_{k}^{\dagger}e^{-i\omega_{d,k}t}-a_{k}e^{i\omega_{d,k}t}\right).
\end{align}
Here $H_{\mathrm{0}}$ describes four eigenmodes of the system, including
two optical modes ($a_{j},~a_{k}$) and two mechanical modes ($b_{j-k},~m$);
$H_{\mathrm{SBS}}$ describes the triple-resonant stimulated Brillouin
scattering between the high-frequency traveling phonon mode and two
optical modes, and $H_{\mathrm{BM}}$ is the dispersive optomechanical
coupling between the breath mechanical mode and optical mode; $H_{\mathrm{D}}$
represents the external laser driving onto the optical modes. $a_{j(k)}$
is the annihilation operator for optical mode $j\left(k\right)$ with
the frequency $\omega_{j\left(k\right)}$, $b_{j-k}$ is the annihilation
operator of the Brillouin mechanical mode with the frequency $\omega_{b}$,
$m$ is the annihilation operator of the breath mechanical mode with
the frequency $\omega_{m}$, $g_{b}$ is the triple-resonant coupling
strength of the stimulated Brillouin scattering, $g_{m,j\left(k\right)}$
is the dispersive coupling rate, $\epsilon_{d,j\left(k\right)}$ is
the driving field of the optical mode $a_{j(k)}$ with a frequency
$\omega_{d,j\left(k\right)}$, and $\kappa_{j\left(k\right),ex}$
is the external coupling rate. For the stimulated Brillouin scattering,
we need both the energy and momentum conservation, so $\omega_{j}=\omega_{\kappa}+\omega_{b}$
is also satisfied.

For a very weak coupling rate $g_{0},~g_{j},~g_{k}\ll\omega_{b},~\omega_{m},~\kappa_{j\left(k\right)}$,
where $\kappa_{j\left(k\right)}=\kappa_{j\left(k\right),o}+\kappa_{j\left(k\right),ex}$
is the total optical energy decay rate, the optomechanical interactions
can be enhanced by the control laser $\epsilon_{d,j\left(k\right)}$.
When the duration of the control pulse $\tau_{d}\gg1/\kappa_{j\left(k\right)}$,
the intracavity control fields can be treated classically as:
\begin{align}
\alpha_{j\left(k\right)} & =\frac{\sqrt{\kappa_{j\left(k\right),ex}}\epsilon_{d,j\left(k\right)}}{\kappa_{j\left(k\right)}/2+i\Delta_{j\left(k\right)}},
\end{align}
where $\Delta_{j\left(k\right)}=\omega_{j\left(k\right)}-\omega_{d,j\left(k\right)}$.
In the interaction picture $H_{\mathrm{0}}=\omega_{d,j}a_{j}^{\dagger}a_{j}+\omega_{d,k}a_{k}^{\dagger}a_{k}+\left(\omega_{d,j}-\omega_{d,k}\right)b_{j-k}^{\dagger}b_{j-k}$,
the Hamiltonian of the system can be linearized as
\begin{align}
H_{\mathrm{lin}}\approx & \Delta_{j}a_{j}^{\dagger}a_{j}+\Delta_{k}a_{k}^{\dagger}a_{k}+\Delta_{b}b_{j-k}^{\dagger}b_{j-k}+\omega_{m}m^{\dagger}m\nonumber \\
 & +G_{b,j}a_{k}^{\dagger}b_{j-k}^{\dagger}+G_{b,k}a_{j}^{\dagger}b_{j-k}+\mathrm{H.c.}\nonumber \\
 & +\left(G_{m,j}a_{j}^{\dagger}+G_{m,k}a_{k}^{\dagger}+\mathrm{H.c.}\right)\left(m^{\dagger}+m\right),
\end{align}
where $\Delta_{b}=\omega_{b}+\omega_{d,k}-\omega_{d,j}$, $G_{b,j\left(k\right)}=g_{b}\alpha_{j\left(\kappa\right)}$,
and $G_{m,j\left(k\right)}=g_{m,j\left(k\right)}\alpha_{j\left(\kappa\right)}$.

\section{Dissipation engineering}

Firstly, we only consider the stimulated Brillouin scattering, as
the mechanical mode $b_{j-k}$ is used as an ancillary to engineer
the optical mode. For the stimulated Brillouin scattering, the triple-resonant
condition is required \citep{dong2015brillouin,zhang2017optomechanical}.
For simplification, we assume that one driving laser is on-resonant
with the optical mode, i.e. $\Delta_{j}=0$ or $\Delta_{k}=0$. By
treating the laser pump on $a_{j}$ as a classical field, we can write
the effective Hamiltonian as $H_{\mathrm{SBS}}=G_{b,j}a_{k}^{\dagger}b_{j-k}^{\dagger}+G_{b,j}^{\ast}a_{k}b_{j-k}$.
Such a stimulated Brillouin scattering is a Stokes process, which
generates photon-phonon pairs into the system, thus eventually induces
gain to the system and reduces the linewidth of the optical mode $a_{k}$.

To observe the effect of dissipation engineering on the optical linewidth,
a weak probe laser is sent into the optical mode $a_{k}$, and the
Langevin equations can be written as follows
\begin{eqnarray}
\frac{d}{dt}a_{k} & = & -\left(\frac{\kappa_{k}}{2}-i\delta\right)a_{k}-iG_{b,j}b_{j-k}^{\dagger}+\sqrt{\kappa_{k,ex}}\epsilon_{p,k},\\
\frac{d}{dt}b_{j-k}^{\dagger} & = & -\left(\frac{\gamma_{b}}{2}-i\delta\right)b_{j-k}^{\dagger}+iG_{b,j}^{\ast}a_{k},
\end{eqnarray}
where $\delta=\omega_{p}-\omega_{k}$, $\gamma_{b}$ is the Brillouin
mechanical decay rate, and $\epsilon_{p,k}$ is the probe laser with
the frequency $\omega_{p}$. When the system is at the steady state,
which requires that the optomechanical coupling is below the lasing
threshold \citep{grudinin2010phonon}, we have $a_{k}\left(\delta\right)=\sqrt{\kappa_{k,ex}}\epsilon_{p,k}/\left(\kappa_{\mathrm{eff}}/2-i\delta_{\mathrm{eff}}\right)$,
where the effective dissipation rate and detuning of the optical mode
can be written as
\begin{align}
\kappa_{\mathrm{eff}} & =\kappa_{k}-\frac{\left|G_{b,j}\right|^{2}\gamma_{b}}{\left(\frac{\gamma_{b}}{2}\right)^{2}+\delta^{2}},\\
\delta_{\mathrm{eff}} & =\delta+\frac{\left|G_{b,j}\right|^{2}\delta}{\left(\frac{\gamma_{b}}{2}\right)^{2}+\delta^{2}}.
\end{align}
If these parameters satisfy $\delta,~\left|G_{b,j}\right|,~\kappa_{k}\ll\gamma_{b}$,
we can obtain the effective decay rate $\kappa_{\mathrm{eff}}\approx\kappa_{k}-4\left|G_{b,j}\right|^{2}/\gamma_{b}$
and the effective detuning $\delta_{\mathrm{eff}}\approx\delta$.

If we consider the driving laser resonant with the optical mode $a_{k}$,
the stimulated Brillouin scattering is an anti-Stokes process, and
the effective Hamiltonian can be written as $H_{\mathrm{SBS}}=G_{b,k}a_{j}^{\dagger}b_{j-k}+G_{b,k}^{\ast}a_{j}b_{j-k}^{\dagger}$.
As opposed to the Stokes process, the particle number conserves and
the coupling would introduce extra loss channel to the optical mode.
Similarly, then the efffective decay rate is derived as $\kappa_{\mathrm{eff}}\approx\kappa_{j}+4\left|G_{b,k}\right|^{2}/\gamma_{b}$,
which confirms that linewidth of the optical mode $a_{j}$ is broadened
due to $b$. In practical experiments, we can detect the intracavity
power to show the variation of the optical mode linewidth. For the
sake of illustration, we can normalize the intracavity power as
\begin{equation}
P_{a}\left(\delta\right)=\left|\frac{\kappa_{\mathrm{eff}}}{2\sqrt{\kappa_{ex}}\epsilon_{p}}a\right|^{2},
\end{equation}
where we have omitted the subscript of some parameters for convenience.

\begin{figure}[htbp]
\center \includegraphics[width=1\columnwidth]{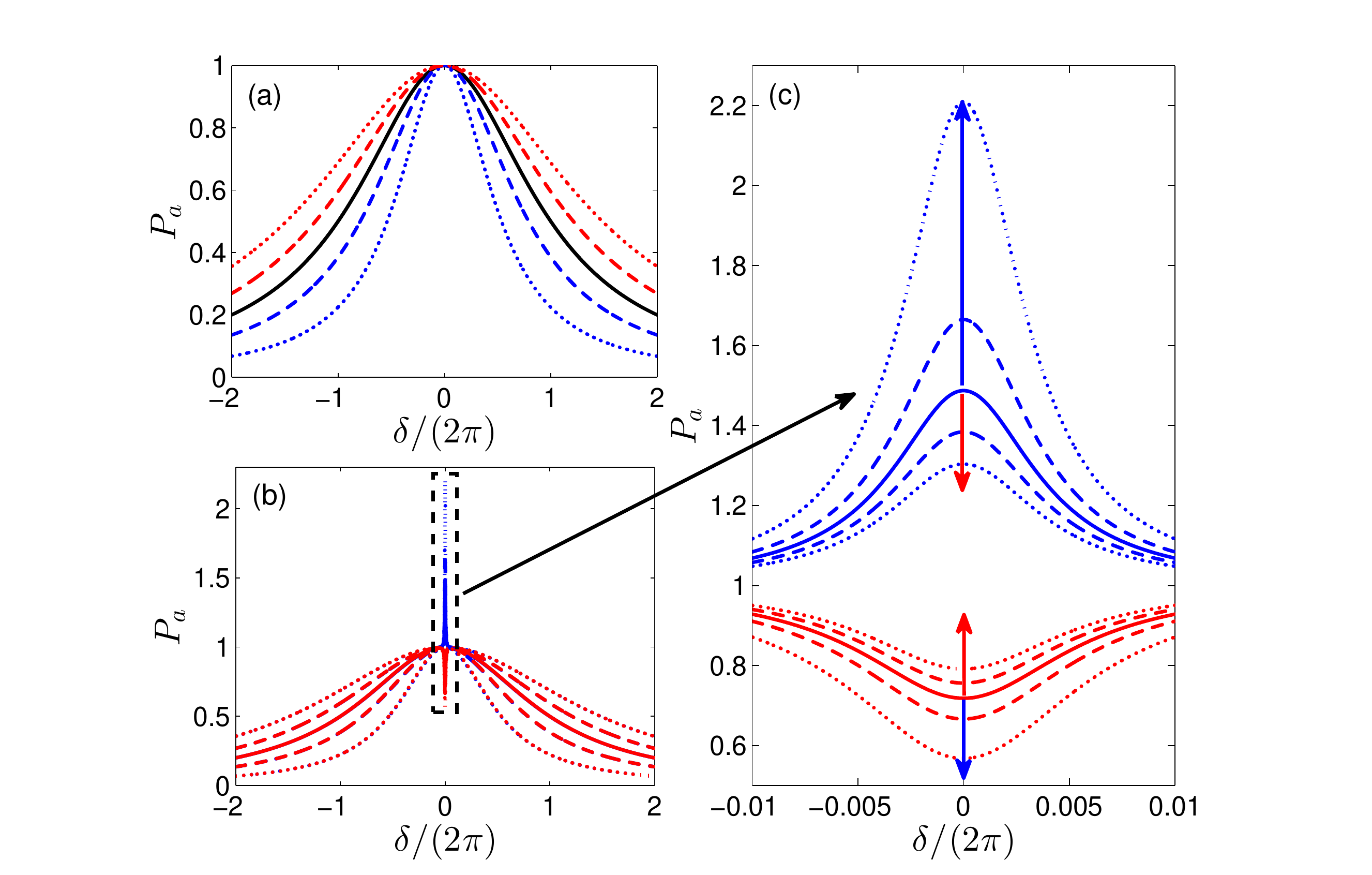} \caption{(Color online) Intracavity power $P_{a}$ of the optical mode $a$.
(a) Engineering the linewidth of the optical mode via only stimulated
Brillouin scattering. The intracavity power spectra for different
control powers $G_{b}/\left(2\pi\right)=0$ (solid line), $2$ (dashed
line), and $3$ (dotted line) $\mathrm{MHz}$, where the blue and
red lines are corresponding to $\Delta_{j}=0$ and $\Delta_{k}=0$,
respectively. (b) Optomechanical induced transparency and amplification
are affected by the dissipation engineering, and (c) is the enlarged
view of the (b) around the detuning $\delta/\left(2\pi\right)\in\left[-0.01,0.01\right]$
$\mathrm{MHz}$. The intracavity power spectra for the red detuning
$\Delta_{j\left(k\right)}=\omega_{m}$ and the blue detuning $\Delta_{j\left(k\right)}=-\omega_{m}$
with the control power $G_{m,j\left(k\right)}/2\pi=0.03$ $\mathrm{MHz}$,
where the blue and red arrows are corresponding to $\Delta_{j}=0$
(Stokes process ) and $\Delta_{k}=0$ (anti-Stokes process ), respectively.
Other parameters are $\kappa_{j}/2\pi=2$ $\mathrm{MHz}$, $\kappa_{k}/2\pi=2$
$\mathrm{MHz}$, $\gamma_{b}/2\pi=40$ $\mathrm{MHz}$, $\omega_{b}/2\pi=10$
$\mathrm{GHz}$, $\gamma_{m}/2\pi=10$ $\mathrm{kHz}$ and $\omega_{m}/2\pi=100$
$\mathrm{MHz}$.}

\label{Fig2}
\end{figure}
In Fig.~\ref{Fig2}(a), we plot the intracavity power $P_{a}$ as
a function of the detuning $\delta$. The blue and red lines correspond
to the case that the control laser is added into the optical mode
$a_{j}$ and $a_{k}$, respectively, and the black line is the result
without the control laser ($G_{b}=0$) for comparison. The dashed
and dotted lines are corresponding to the effective coupling $G_{b}/\left(2\pi\right)=2$
and $3$ $\mathrm{MHz}$, respectively. When $\Delta_{j}=0$, the
blue lines show that the effective linewidth of the optical mode is
reduced with the increasing of the control powers, and the intracavity
power spectra agree with the theoretical result $\kappa_{\mathrm{eff}}=\kappa_{k}-4\left|G_{b,j}\right|^{2}/\gamma_{b}$.
For $\Delta_{k}=0$, the related results are plotted by the red lines,
which show that the broadened linewidth of the optical mode as $\kappa_{\mathrm{eff}}=\kappa_{j}+4\left|G_{b,k}\right|^{2}/\gamma_{b}$.
Based on the analytical derivations and numerical calculation above,
we conclude that the effective linewidth of the optical mode can be
effectively reduced or broadened by the dissipation engineering.

Now we consider the breath mechanical mode $m$ for the radiation
pressure optomechanical coupling, where the OMIT or OMIA can be observed
due to another drive laser that induces coherent photon-phonon interaction.
From Fig.~\ref{Fig2}(a), we know that the effective linewidth of
the optical mode can be modulated by the stimulated Brillouin scattering,
which indicates that the OMIT and OMIA can be effectively controlled
by the dissipation engineering. For the radiation pressure optomechanical
coupling, the drive can be either red or blue detuning $\Delta_{j\left(k\right)}=\pm\omega_{m}$.
Combining the stimulated Brillouin scattering and optomechanical couplings,
we can divide possible experimental configurations into four cases:
(a) anti-Stokes process and red detuning: $\Delta_{k}=0,~\Delta_{j}=\omega_{m}$;
(b) anti-Stokes process and blue detuning: $\Delta_{k}=0,~\Delta_{j}=-\omega_{m}$;
(c) Stokes process and red detuning: $\Delta_{j}=0,~\Delta_{k}=\omega_{m}$;
(d) Stokes process and blue detuning: $\Delta_{j}=0,~\Delta_{k}=-\omega_{m}$.
To illustrate different situations intuitively, we add a weak laser
to probe the optical mode $a$. Firstly, we consider the case (a)
and the corresponding Hamiltonian of the system can be written as
\begin{eqnarray}
H & = & -\delta a^{\dagger}a-\delta b^{\dagger}b-\delta m^{\dagger}m+\left(G_{b}a^{\dagger}b+G_{m}a^{\dagger}m+\mathrm{H.C.}\right)\nonumber \\
 &  & +i\sqrt{\kappa_{ex}}\epsilon_{p}\left(a^{\dagger}-a\right),
\end{eqnarray}
where $\delta=\omega_{p}-\omega_{j}$. We have omitted the subscript
of some parameters for convenience and neglected other terms with
the rotating wave approximation $\omega_{m}\gg\kappa_{j\left(k\right)},\gamma_{m}$,
where $\gamma_{m}$ is the breath mechanical decay rate. The steady
state solution reads
\begin{align}
a\left(\delta\right) & =\frac{\sqrt{\kappa_{ex}}\epsilon_{p}}{\frac{\kappa}{2}-i\delta+\frac{G_{b}^{2}}{\gamma_{b}/2-i\delta}+\frac{G_{m}^{2}}{\gamma_{m}/2-i\delta}}.
\end{align}
For the convenience, we have assumed that $G_{b}$ and $G_{m}$ are
real numbers. When we consider other cases (b,c,d), the steady state
solution has a similar form to the above equation and we only change
the couplings: $G_{b}^{2}\rightarrow-G_{b}^{2}$ for the Stokes process;
$G_{m}^{2}\rightarrow-G_{m}^{2}$ for the blue detuning.

In Figs.~\ref{Fig2}(b) and (c), we discuss the dissipation engineering
on both the OMIT and OMIA, and we plot the normalized intracavity
power $P_{a}$ as the function of the detuning between the probe laser
and the optical cavity. When the probe drive is largely detuned from
the mechanical resonance, the effect due to radiation pressure optomechanical
coupling is almost not observable, where the Fig.~\ref{Fig2}(b)
is almost the same as Fig.~\ref{Fig2}(a). Figure~\ref{Fig2}(c)
is the magnification of Fig.~\ref{Fig2}(b) around the detuning $\delta/\left(2\pi\right)\in\left[-0.01,0.01\right]$
$\mathrm{MHz}$. The arrows mean that the OMIT and OMIA is enhanced
(the blue arrow) or suppressed (the red arrow) with the increasing
of dissipation engineering ($G_{b}$). Here the parameters that we
choose satisfy $\gamma_{b}\gg\kappa_{k\left(j\right)},\gamma_{m}$,
and we only consider the detuning $\left|\delta\right|\ll\gamma_{b}$.
These phenomena can be described by the approximate emission power
spectra
\begin{equation}
P_{a}\approx\left|\frac{1}{1-2i\delta/\kappa_{\mathrm{eff}}\pm\frac{C}{1-2i\delta/\gamma_{m}}}\right|^{2},
\end{equation}
where the cooperativity $C=4\left|G_{m}\right|^{2}/\left(\kappa_{\mathrm{eff}}\gamma_{m}\right)$
and $\pm$ for the blue and red detuning. For a fixed $G_{m}$, it
is obvious that the OMIT and OMIA are controlled by the effective
linewidth of the optical mode $\kappa_{\mathrm{eff}}=\kappa\pm4\left|G_{b}\right|^{2}/\gamma_{b}$.

From Fig.~\ref{Fig2}, it is demonstrated that the dissipation engineering
could modulate the linewidth of optical mode, which can be used to
control the OMIT and OMIA. Especially, it can also make the unresolvable
sideband in the optomechanical system, i.e. $\omega_{m}<\kappa_{\mathrm{eff}}$,
to be resolvable by reducing the linewidth of the optical mode.

\section{Entanglement and cooling}

Next, we discuss the effect of the dissipation engineering on the
entanglement of radiation pressure optomechanical system and cooling
of the mechanical mode $m$. It is known that the thermal noise is
bad for the entanglement \citep{tian2013robust} and cooling, so we
only consider the anti-Stokes process as the dissipation engineering
to cool the system. For the stimulated Brillouin scattering, we consider
the driving laser is resonant with the optical mode $a_{k}$. The
Hamiltonian of the system can be written as
\begin{align}
H= & \Delta_{j}a_{j}^{\dagger}a_{j}+\Delta_{b}b_{j-k}^{\dagger}b_{j-k}+\omega_{m}m^{\dagger}m\nonumber \\
 & +\left[G_{b,k}a_{j}^{\dagger}b_{j-k}+G_{m,j}a_{j}^{\dagger}\left(m^{\dagger}+m\right)+\mathrm{H.C.}\right].
\end{align}
We define $R=\left(q_{a},p_{a},q_{b},p_{b},q_{m},p_{m}\right)^{T}$,
where $q_{o}=\left(o^{\dagger}+o\right)/\sqrt{2}$ and $p_{o}=i\left(o^{\dagger}-o\right)/\sqrt{2}$.
The Langevin equations can be written as
\begin{equation}
\frac{d}{dt}R=MR+R_{in},
\end{equation}
where
\begin{equation}
M=\left[\begin{array}{cccccc}
-\frac{\kappa_{k}}{2} & \Delta_{j} & 0 & G_{b,k} & 0 & 0\\
-\Delta_{j} & -\frac{\kappa_{k}}{2} & -G_{b,k} & 0 & -2G_{m,j} & 0\\
0 & G_{b,k} & -\frac{\gamma_{b}}{2} & \Delta_{b} & 0 & 0\\
-G_{b,k} & 0 & -\Delta_{b} & -\frac{\gamma_{b}}{2} & 0 & 0\\
0 & 0 & 0 & 0 & -\frac{\gamma_{m}}{2} & \omega_{m}\\
-2G_{m,j} & 0 & 0 & 0 & -\omega_{m} & -\frac{\gamma_{m}}{2}
\end{array}\right],
\end{equation}
and the thermal noise can be written as $R_{in}=\left(\sqrt{\kappa_{j}}q_{a}^{in},\sqrt{\kappa_{j}}p_{a}^{in},\sqrt{\gamma_{b}}q_{b}^{in},\sqrt{\gamma_{b}}p_{b}^{in},\sqrt{\gamma_{m}}q_{m}^{in},\sqrt{\gamma_{m}}p_{m}^{in}\right)^{T}$.
Here we have assumed that the $G_{b,k}$ and $G_{m,j}$ are the real
numbers. The fact that the dynamics of the system is governed by a
linearized Hamiltonian ensures that the evolved states are Gaussian
states whose information-related properties \citep{weedbrook2012gaussian}
are fully represented by the 6 \texttimes{} 6 covariance matrix with
entries defined as
\begin{equation}
V_{i,j}=\left\langle R_{i}R_{j}+R_{j}R_{i}\right\rangle /2.
\end{equation}
The equation of motion corresponding to the covariance matrix can
be written as follows
\begin{equation}
\frac{d}{dt}V=MV+VM^{T}+D,
\end{equation}
where $D=\mathrm{Diag}\left(\frac{\kappa_{k}}{2},\frac{\kappa_{k}}{2},\frac{\gamma_{b}}{2},\frac{\gamma_{b}}{2},\frac{\gamma_{m}\left(2n_{m}+1\right)}{2},\frac{\gamma_{m}\left(2n_{m}+1\right)}{2}\right)$.
For $\omega_{b}\gg\omega_{m}$, we have neglected the thermal noise
of the mode $b$.

\subsection{Entanglement}

The entanglement between optical mode and mechanical mode is generated
by the interaction term $G_{m,j}a_{j}^{\dagger}m^{\dagger}+\mathrm{H.c.}$,
so we choose the detuning $\Delta_{j}=\Delta_{b}=-\omega_{m}$. The
continuous variable entanglement of two modes can be calculated by
the logarithmic negativity $E_{N}$ \citep{plenio2005logarithmic}.
This quantity is a rigorous entanglement monotone, and is zero for
separable states. For the two-mode Gaussian states, it can be calculated
using the expression \citep{wang2013reservoir}\textcolor{red}{{} }
\begin{equation}
E_{N}=\max\left[0,-\ln\left(2\eta\right)\right],
\end{equation}
where
\begin{equation}
\eta=\frac{1}{\sqrt{2}}\sqrt{\Sigma-\sqrt{\Sigma^{2}-4\det V}},
\end{equation}
and $\Sigma=\det V_{1}+\det V_{2}-\det V_{3}$. The matrix $V_{1}$,
$V_{2}$ and $V_{3}$ are 2$\times$2 matrices related to the covariance
matrix as
\begin{equation}
V=\left[\begin{array}{cc}
V_{1} & V_{3}\\
V_{3}^{T} & V_{2}
\end{array}\right].
\end{equation}

\begin{figure}[htbp]
\center \includegraphics[width=1\columnwidth]{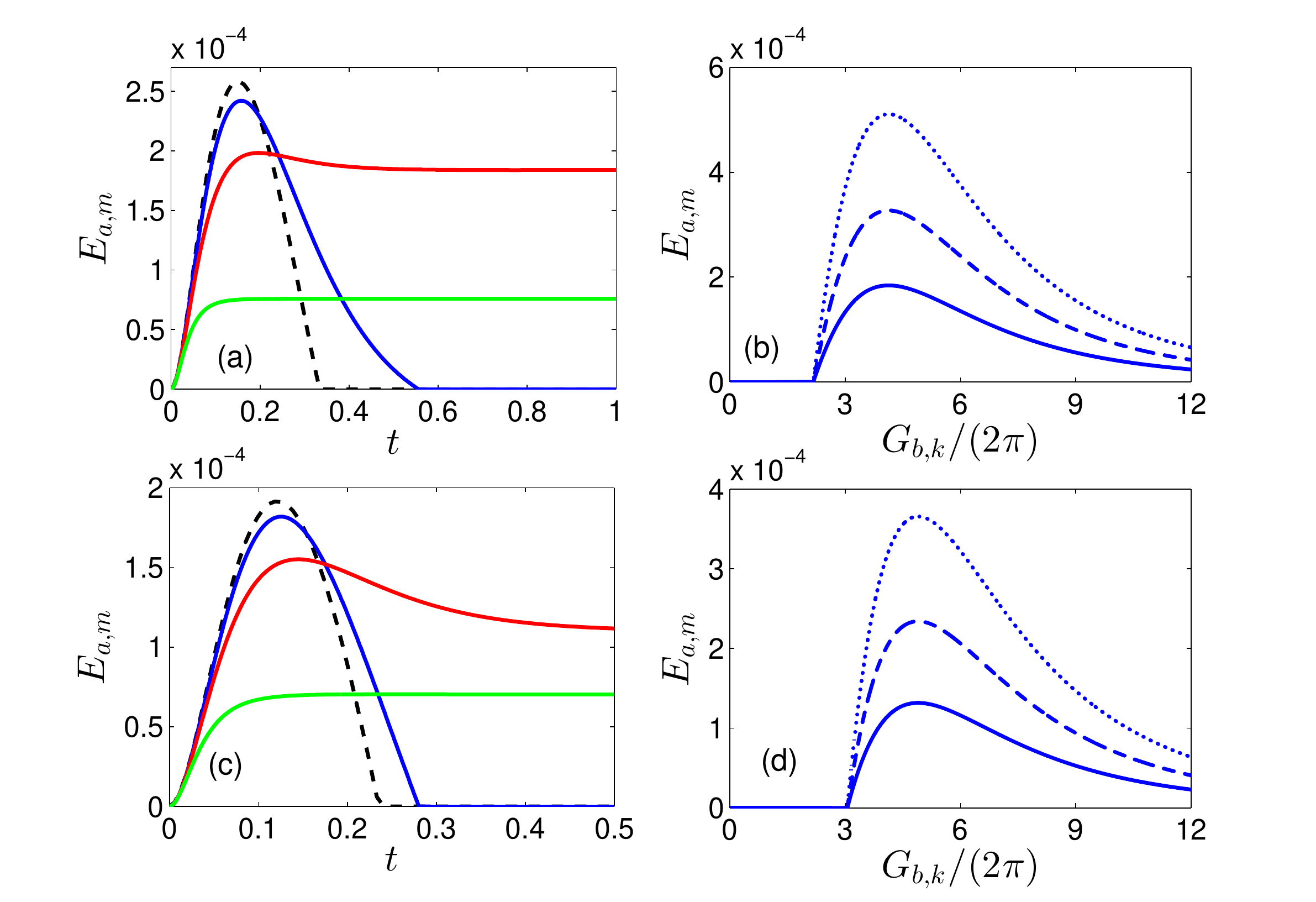} \caption{(Color online) Logarithmic negativity $E_{am}$ affected by the dissipation
engineering. (a) The logarithmic negativity as a funtion of the time
$t$ with $G_{m,j}/\left(2\pi\right)=0.03$ MHz for $G_{b,k}/\left(2\pi\right)=0$
(black dashed line), $2$ (blue line), $4$ (red line), and $8$ (green
line) MHz. (b) The logarithmic negativity versus the dissipation engineering
$G_{b,k}$ for $G_{m,j}/\left(2\pi\right)=0.03$ (solid line), $0.04$
(dashed line), $0.05$ (dotted line) MHz. (a) and (b) are the results
with the thermal noise $n_{m}=250$; (c) and (d) are the condition
with $n_{m}=300$. Other parameters are the same as Fig. \ref{Fig2}. }

\label{Fig3}
\end{figure}
Here we only concern the entanglement of radiation pressure optomechanical
system, so the covariance matrix $V$ is about the modes $a,~m$.
In Fig.~\ref{Fig3}, we study the effect of dissipation engineering
on the entanglement. The dynamical logarithmic negativity $E_{am}$
with dissipation engineering is shown in Fig.~\ref{Fig3}(a). The
black dashed line is the result without the dissipation engineering,
and the $E_{am}$ reaches a maximum at an appropriate evolution time.
When the system approaches the steady-state, the entanglement disappears
with $E_{am}=0$ for the thermal noise $n_{m}$. The solid lines show
the dissipation engineering on the entanglement, where $G_{b,k}/\left(2\pi\right)=2$
(blue line) and $4$ (red line) MHz, which mean that the steady logarithmic
negativity $E_{am}$ becomes bigger with the increasing of the dissipation
engineering $G_{b,k}$. However, for stronger dissipation engineering
$G_{b,k}/\left(2\pi\right)=8$ (green line) MHz, it will reduce the
steady logarithmic negativity, which means that there is the optimal
engineering for the logarithmic negativity.

In Fig.~\ref{Fig3}(b), we plot the optimal engineering for different
couplings $G_{m,j}/\left(2\pi\right)=0.03$ (solid line), $0.04$
(dashed line), $0.05$ (dotted line) MHz, and the steady logarithmic
negativity $E_{am}$ has a peak. Because of the thermal noise $n_{m}$,
there is a phase transition point for the entanglement growing out
of nothing. For more thermal noise $n_{m}=300$, the Figs.~\ref{Fig3}(c)
and (d) show that the logarithmic negativity $E_{am}$ becomes smaller
and it needs more dissipation engineering to recover the entanglement.

\begin{figure}[htbp]
\center \includegraphics[width=1\columnwidth]{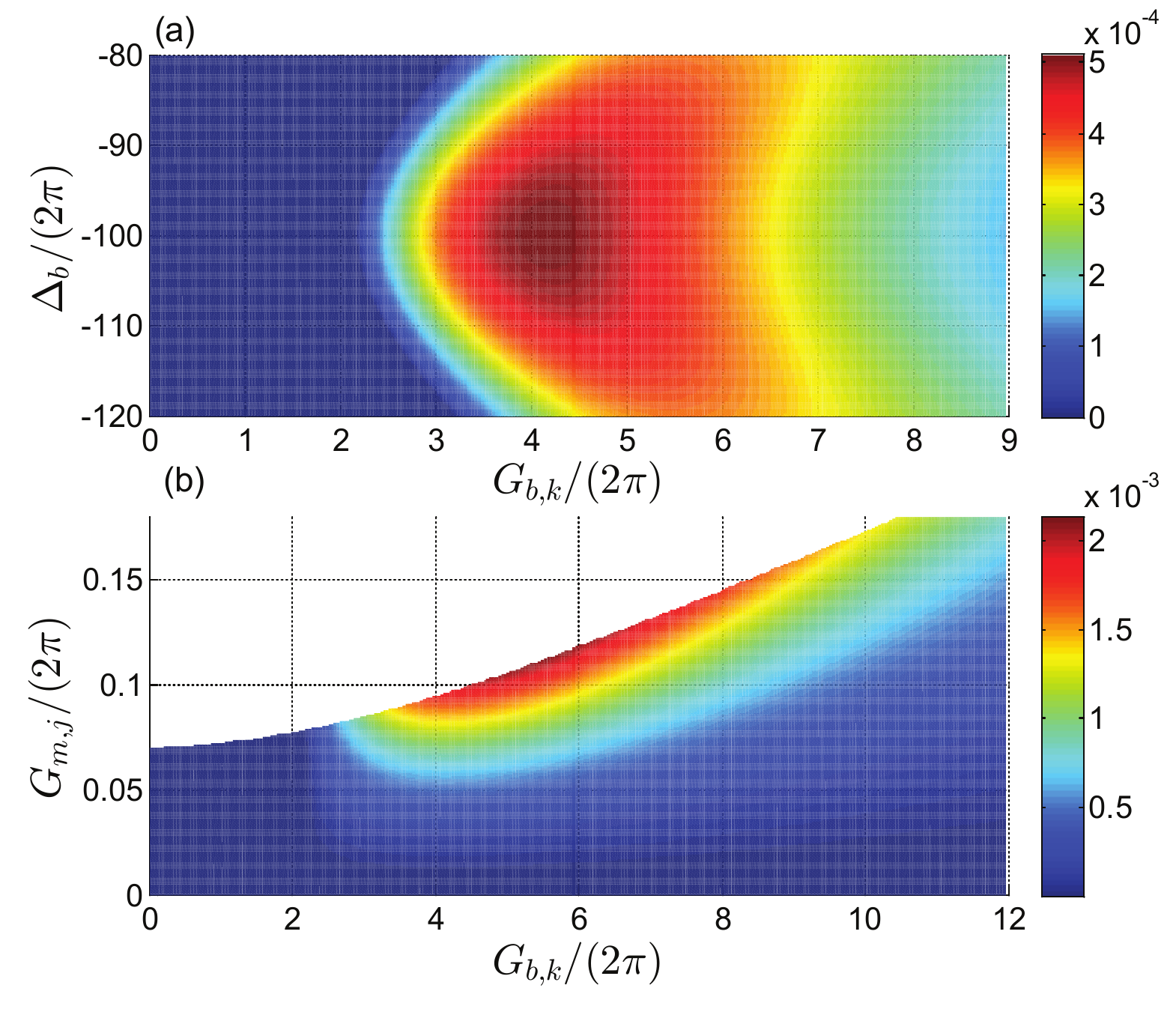} \caption{(Color online) (a) Logarithmic negativity $E_{am}$ as a function
of both $\Delta_{b}$ and $G_{b,k}$ with $G_{m,j}/\left(2\pi\right)=0.05$
MHz. (b) Logarithmic negativity $E_{am}$ for both $G_{m,j}$ and
$G_{b,k}$ with $\Delta_{b}=-\omega_{m}$. Here we choose the $n_{m}=250$.
Other parameters are the same as Fig.~\ref{Fig2}.}

\label{Fig4}
\end{figure}
For dissipation engineering, we also consider the effect of the detuning
$\Delta_{b}$ on the stationary intracavity entanglement, which is
shown in Fig.~\ref{Fig4}(a). Near the resonance point $\Delta_{b}=-\omega_{m}$,
we obtain the maximal $E_{am}$, which is due to the most effective
coupling at the resonance point. With the increas of the coupling
$G_{m,j}$, we can always find an optimal $G_{b,k}$ for the stationary
intracavity entanglement, as shown in Fig.~\ref{Fig4}(b). However,
for the large $G_{m,j}$, the eigenvalues of $M$ have a positive
real part, which makes the system unstable. We can predicate stability
conditions by the well-known Routh-Hurwitz criteria \citep{dejesus1987routh},
which is labeled by the blank in Fig.~\ref{Fig4}(b). The maximal
$E_{am}$ is obtained at the edge of the unstable region, and the
stable entanglement becomes weaker for the larger $G_{m,j}$.

From Figs.~\ref{Fig3} and \ref{Fig4}, we conclude that the noise-overwhelmed
steady entanglement between the optical mode and mechanical mode can
be restored by the dissipation engineering, which is robust to the
thermal noise of the mechanical mode. It is due to the anti-Stokes
process from the stimulated Brillouin scattering to cool the optomechanical
system. For the strong dissipation engineering, the entanglement abates
for the strong-nonlinearity.

\subsection{Cooling}

The dissipation engineering can be also used to enhance the cooling
of the mechanical mode $m$. From the covariance matrix, the final
steady thermal occupation of the mode $m$ is calculated by
\begin{align}
\left\langle m^{\dagger}m\right\rangle  & =n_{f}=\frac{V_{5,5}+V_{6,6}-1}{2}.
\end{align}

\begin{figure}[htbp]
\center \includegraphics[width=1\columnwidth]{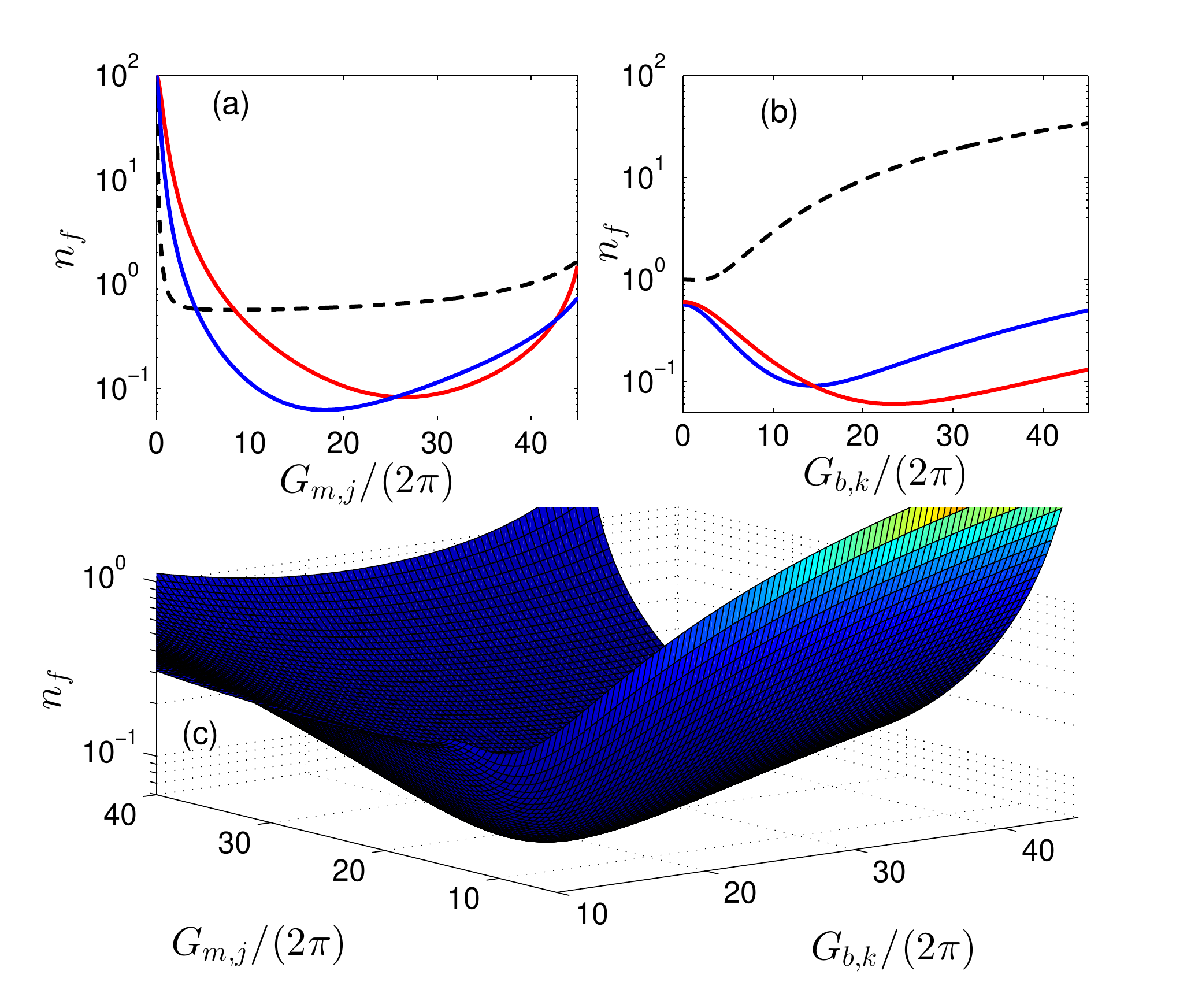} \caption{(Color online) (a) The thermal occupation as a function of $G_{m,j}$
for $G_{b,k}/\left(2\pi\right)=0$ (black dashed line), $20$ (blue
line) and $40$ (red line) MHz. (b) The thermal occupation as a function
of $G_{b,k}$ for $G_{m,j}/\left(2\pi\right)=1$ (black dashed line),
$10$ (blue line) and $20$ (red line) MHz. (c) Cooling effect for
both $G_{m,j}$ and $G_{b,k}$ . Here we choose the $n_{m}=100$.
Other parameters are the same as Fig.~\ref{Fig2}.}

\label{Fig5}
\end{figure}
The cooling is from the interaction term $G_{m,j}a_{j}^{\dagger}m+\mathrm{H.c.}$~\citep{aspelmeyer2014cavity},
so we choose the detuning $\Delta_{j}=\Delta_{b}=\omega_{m}$. In
Fig.~\ref{Fig5}, we discuss the cooling effect of dissipation engineering.
We plot the thermal occupation $n_{f}$ as a function of $G_{m,j}$
in Fig.~\ref{Fig5}(a). The black dashed line is the result without
the dissipation engineering, and the thermal noise reduces quickly
with the increasing of $G_{m,j}$. And then it goes up slowly, which
is corresponding to the term $G_{m,j}a_{j}^{\dagger}m^{\dagger}+\mathrm{H.c.}$,
where the rotating wave approximation is not well satisfied for very
large $G_{m,j}$. When $G_{b,k}/\left(2\pi\right)=20$~MHz, we plot
the blue line, and the cooling effect becomes worse than no dissipation
engineering in the weak coupling area, where $G_{m,j}/\left(2\pi\right)\leq1$~MHz.
This can be explained by the classical concept, where the linewidth
of the optical mode $a_{k}$ can be broadened as $\kappa_{\mathrm{eff}}=\kappa_{k}+4\left|G_{b,k}\right|^{2}/\gamma_{b}$
for the coupling $G_{b,k}a_{j}^{\dagger}b_{j-k}+\mathrm{H.c.}$, and
the final thermal occupation in equilibrium is calculated by the rate
equation as
\begin{equation}
n_{f}=\frac{\gamma_{m}n_{m}}{\gamma_{m}+\frac{4G_{m,j}^{2}}{\kappa_{\mathrm{eff}}}}.
\end{equation}
However, the cooling is greatly enhanced for the strong coupling area
with $G_{m,j}/\left(2\pi\right)\geq10$~MHz comparing with no dissipation
engineering, and the classical concept no longer holds. For a very
strong dissipation engineering coupling strength $G_{b,j}/\left(2\pi\right)=40$~MHz,
the red line shows that the cooling effect becomes bad again, and
it is due to the frequency shift of the modes for strong dissipation
engineering. The modes will become the supermodes, which mean $a_{j}\Rightarrow\left(A+B\right)/\sqrt{2}$
and $b\Rightarrow\left(B-A\right)/\sqrt{2}$, and the corresponding
frequencies are transformed as $\omega_{A}=\omega_{m}-\left|G_{b,k}\right|$
and $\omega_{B}=\omega_{m}+\left|G_{b,k}\right|$. It is obvious that
the effective term $G_{m,j}A^{\dagger}m^{\dagger}+\mathrm{H.c.}$
can not be absolutely eliminated by the rotating wave approximation
for strong dissipation engineering. Therefore we can obtain the optimal
optomechanical cooling with the increasing of dissipation engineering
coupling strength.

For the weak coupling, the dissipation engineering still has a bad
effect on the cooling, which is observed by the black dashed line
in Fig.~\ref{Fig5}(b). The blue and red lines show that there is
a minimal value of $n_{f}$ by the dissipation engineering for the
strong coupling. Taking the coupling $G_{m,j}$ and the dissipation
engineering $G_{b,k}$ into consideration, we plot the Fig.~\ref{Fig5}(c).
It is obviously shown that it can improve the cooling effect by an
order of magnitude compared to the case without dissipation engineering.

\begin{figure}[htbp]
\center \includegraphics[width=1\columnwidth]{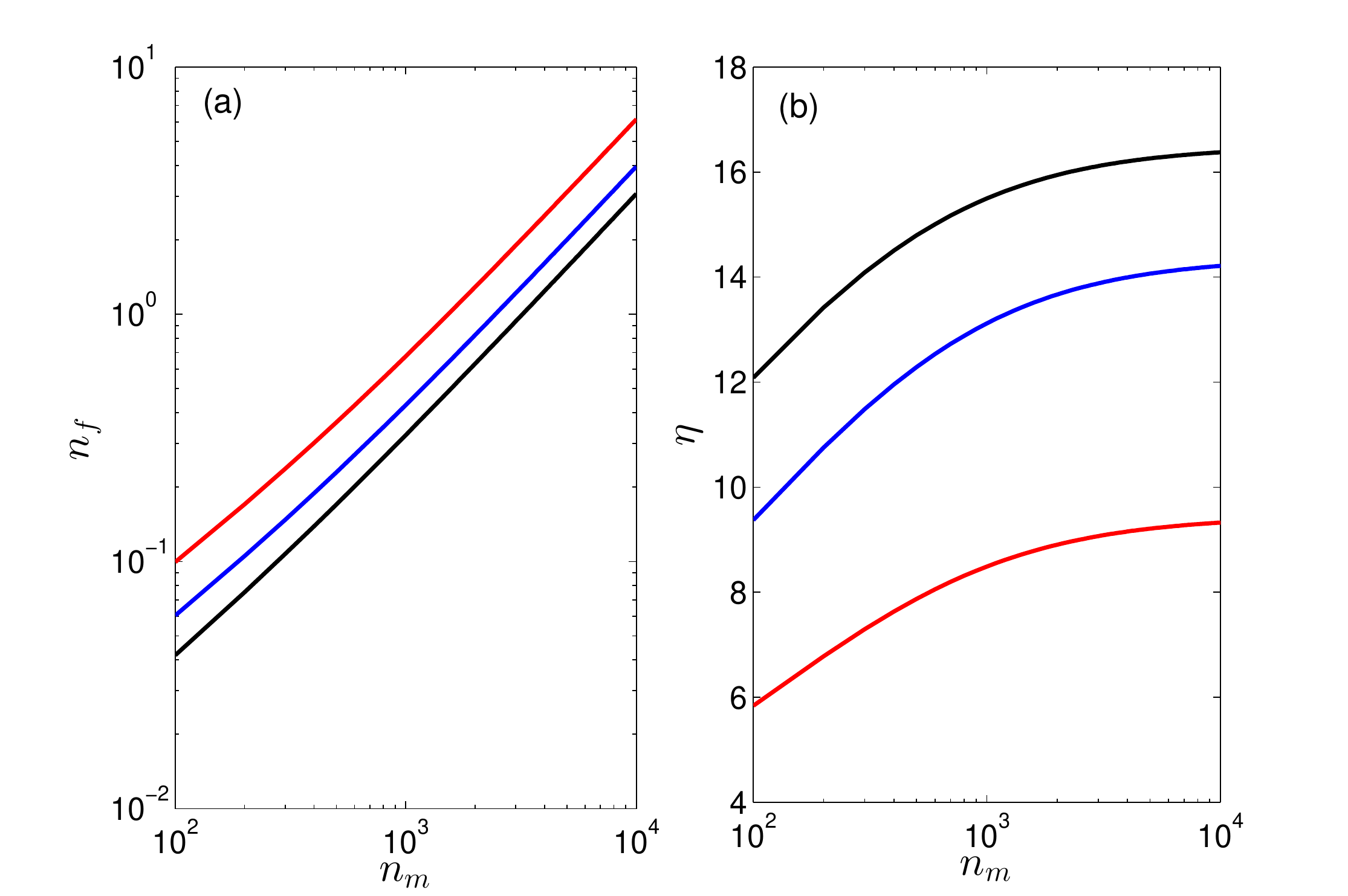} \caption{(Color online) Thermal occupation $n_{f}$ (a) and the ratio (b) as
a function of $n_{m}$ for $\omega_{m}/\left(2\pi\right)=50$ (red
line), $100$ (blue line) and $200$ (black line) MHz. Other parameters
are the same as Fig.~\ref{Fig2}.}

\label{Fig6}
\end{figure}
For different $n_{m}$ and $\omega_{m}$, we can always obtain the
optimal cooling by tuning the coupling $G_{m,j}$ and the dissipation
engineering coupling strength $G_{b,k}$, as shown in Fig.~\ref{Fig5}(c).
In Fig.~\ref{Fig6}, we discuss the optimal cooling as a function
of $n_{m}$ for $\omega_{m}/\left(2\pi\right)=50$ (red line), $100$
(blue line) and $200$ (black line) MHz. From Fig.~\ref{Fig6}(a),
it is observed that the thermal occupation $n_{f}$ is linear to the
thermal noise $n_{m}$ and the cooling effect is better for the higher
frequency $\omega_{m}$, which is due to the reasonable condition
for the rotating wave approximation. Compared with no dissipation
engineering $G_{b,k}=0$, the ratio $\eta=n_{f\left(G_{b,k}=0\right)}/n_{f}$
is plotted in Fig.~\ref{Fig6}(b), and we observe that the cooling
effect is more obvious for the more thermal noise and the ratio value
eventually stabilizes. For the frequency $\omega_{m}$, the ratio
has a similar phenomenon with the thermal occupation $n_{f}$. In
conclusion, we can improve the cooling effect via dissipation engineering
by about an order of magnitude for the low-frequency mechanical modes.

\section{Conclusion}

We theoretically studied a multimode optomechanical system and proposed
the optomechanical dissipation engineering by an ancilla mechanical
mode. It is demonstrated that the dissipation engineering could enhance
the system performances on the optomechanical induced transparency
and amplification, optomechanical entanglement generation, and the
cooling of the mechanical mode. It is shown that the optomechanical
induced transparency and amplification can be enhanced or suppressed
due to the controlling of the optical linewidth through coupling to
ancilla mechanical mode. In the weak regime of the optomechanical
coupling, the thermal noise from the mechanical mode destroys the
entanglement between the optical mode and mechanical mode. However,
we can recover the entanglement by the anti-Stokes process. In the
strong coupling regime, we also find that the optomechanical cooling
can be greatly enhanced by dissipation engineering. For the entanglement
and cooling, the destruction is from the thermal noise, and the anti-Stokes
process provides a cooling effect, which is beneficial to the optomechanical
system. However, the strong dissipation engineering ultimately has
a bad effect on the multimode optomechanical system because of strong
nonlinearity, therefore we can always find the optimal dissipation
engineering to realize the best entanglement and cooling. The proposed
optomechanical dissipation engineering opens up novel prospects for
the phonon-based quantum information processing and macroscopic quantum
phenomena.

{\em Acknowledgments.} This work was funded by the Key R\&D Program
of China (Grant No. 2016YFA0301303), the National Natural Science
Foundation of China (Grant No.11874342, 11934012, 11922411, 11722436,
11674305 and 11704370), the China Postdoctoral Science Foundation
(No. 2019M652181), and Anhui Initiative in Quantum Information Technologies
(AHY130200).
\nocite{*}
\bibliography{main}

\begin{thebibliography}{59}%
\makeatletter
\providecommand \@ifxundefined [1]{%
 \@ifx{#1\undefined}
}%
\providecommand \@ifnum [1]{%
 \ifnum #1\expandafter \@firstoftwo
 \else \expandafter \@secondoftwo
 \fi
}%
\providecommand \@ifx [1]{%
 \ifx #1\expandafter \@firstoftwo
 \else \expandafter \@secondoftwo
 \fi
}%
\providecommand \natexlab [1]{#1}%
\providecommand \enquote  [1]{``#1''}%
\providecommand \bibnamefont  [1]{#1}%
\providecommand \bibfnamefont [1]{#1}%
\providecommand \citenamefont [1]{#1}%
\providecommand \href@noop [0]{\@secondoftwo}%
\providecommand \href [0]{\begingroup \@sanitize@url \@href}%
\providecommand \@href[1]{\@@startlink{#1}\@@href}%
\providecommand \@@href[1]{\endgroup#1\@@endlink}%
\providecommand \@sanitize@url [0]{\catcode `\\12\catcode `\$12\catcode
  `\&12\catcode `\#12\catcode `\^12\catcode `\_12\catcode `\%12\relax}%
\providecommand \@@startlink[1]{}%
\providecommand \@@endlink[0]{}%
\providecommand \url  [0]{\begingroup\@sanitize@url \@url }%
\providecommand \@url [1]{\endgroup\@href {#1}{\urlprefix }}%
\providecommand \urlprefix  [0]{URL }%
\providecommand \Eprint [0]{\href }%
\providecommand \doibase [0]{http://dx.doi.org/}%
\providecommand \selectlanguage [0]{\@gobble}%
\providecommand \bibinfo  [0]{\@secondoftwo}%
\providecommand \bibfield  [0]{\@secondoftwo}%
\providecommand \translation [1]{[#1]}%
\providecommand \BibitemOpen [0]{}%
\providecommand \bibitemStop [0]{}%
\providecommand \bibitemNoStop [0]{.\EOS\space}%
\providecommand \EOS [0]{\spacefactor3000\relax}%
\providecommand \BibitemShut  [1]{\csname bibitem#1\endcsname}%
\let\auto@bib@innerbib\@empty
\bibitem [{\citenamefont {Aspelmeyer}\ \emph {et~al.}(2012)\citenamefont
  {Aspelmeyer}, \citenamefont {Meystre},\ and\ \citenamefont
  {Schwab}}]{aspelmeyer2012quantum}%
  \BibitemOpen
  \bibfield  {author} {\bibinfo {author} {\bibfnamefont {M.}~\bibnamefont
  {Aspelmeyer}}, \bibinfo {author} {\bibfnamefont {P.}~\bibnamefont {Meystre}},
  \ and\ \bibinfo {author} {\bibfnamefont {K.}~\bibnamefont {Schwab}},\
  }\bibfield  {title} {\enquote {\bibinfo {title} {Quantum optomechanics},}\
  }\href@noop {} {\bibfield  {journal} {\bibinfo  {journal} {Physics Today}\
  }\textbf {\bibinfo {volume} {65}},\ \bibinfo {pages} {29} (\bibinfo {year}
  {2012})}\BibitemShut {NoStop}%
\bibitem [{\citenamefont {Kippenberg}\ and\ \citenamefont
  {Vahala}(2008)}]{kippenberg2008cavity}%
  \BibitemOpen
  \bibfield  {author} {\bibinfo {author} {\bibfnamefont {T.~J.}\ \bibnamefont
  {Kippenberg}}\ and\ \bibinfo {author} {\bibfnamefont {K.~J.}\ \bibnamefont
  {Vahala}},\ }\bibfield  {title} {\enquote {\bibinfo {title} {Cavity
  optomechanics: back-action at the mesoscale},}\ }\href@noop {} {\bibfield
  {journal} {\bibinfo  {journal} {science}\ }\textbf {\bibinfo {volume}
  {321}},\ \bibinfo {pages} {1172} (\bibinfo {year} {2008})}\BibitemShut
  {NoStop}%
\bibitem [{\citenamefont {Van~Thourhout}\ and\ \citenamefont
  {Roels}(2010)}]{van2010optomechanical}%
  \BibitemOpen
  \bibfield  {author} {\bibinfo {author} {\bibfnamefont {D.}~\bibnamefont
  {Van~Thourhout}}\ and\ \bibinfo {author} {\bibfnamefont {J.}~\bibnamefont
  {Roels}},\ }\bibfield  {title} {\enquote {\bibinfo {title} {Optomechanical
  device actuation through the optical gradient force},}\ }\href@noop {}
  {\bibfield  {journal} {\bibinfo  {journal} {Nature Photonics}\ }\textbf
  {\bibinfo {volume} {4}},\ \bibinfo {pages} {211} (\bibinfo {year}
  {2010})}\BibitemShut {NoStop}%
\bibitem [{\citenamefont {Aspelmeyer}\ \emph {et~al.}(2014)\citenamefont
  {Aspelmeyer}, \citenamefont {Kippenberg},\ and\ \citenamefont
  {Marquardt}}]{aspelmeyer2014cavity}%
  \BibitemOpen
  \bibfield  {author} {\bibinfo {author} {\bibfnamefont {M.}~\bibnamefont
  {Aspelmeyer}}, \bibinfo {author} {\bibfnamefont {T.~J.}\ \bibnamefont
  {Kippenberg}}, \ and\ \bibinfo {author} {\bibfnamefont {F.}~\bibnamefont
  {Marquardt}},\ }\bibfield  {title} {\enquote {\bibinfo {title} {Cavity
  optomechanics},}\ }\href@noop {} {\bibfield  {journal} {\bibinfo  {journal}
  {Reviews of Modern Physics}\ }\textbf {\bibinfo {volume} {86}},\ \bibinfo
  {pages} {1391} (\bibinfo {year} {2014})}\BibitemShut {NoStop}%
\bibitem [{\citenamefont {Weis}\ \emph {et~al.}(2010)\citenamefont {Weis},
  \citenamefont {Rivi{\`e}re}, \citenamefont {Del{\'e}glise}, \citenamefont
  {Gavartin}, \citenamefont {Arcizet}, \citenamefont {Schliesser},\ and\
  \citenamefont {Kippenberg}}]{weis2010optomechanically}%
  \BibitemOpen
  \bibfield  {author} {\bibinfo {author} {\bibfnamefont {S.}~\bibnamefont
  {Weis}}, \bibinfo {author} {\bibfnamefont {R.}~\bibnamefont {Rivi{\`e}re}},
  \bibinfo {author} {\bibfnamefont {S.}~\bibnamefont {Del{\'e}glise}}, \bibinfo
  {author} {\bibfnamefont {E.}~\bibnamefont {Gavartin}}, \bibinfo {author}
  {\bibfnamefont {O.}~\bibnamefont {Arcizet}}, \bibinfo {author} {\bibfnamefont
  {A.}~\bibnamefont {Schliesser}}, \ and\ \bibinfo {author} {\bibfnamefont
  {T.~J.}\ \bibnamefont {Kippenberg}},\ }\bibfield  {title} {\enquote {\bibinfo
  {title} {Optomechanically induced transparency},}\ }\href@noop {} {\bibfield
  {journal} {\bibinfo  {journal} {Science}\ }\textbf {\bibinfo {volume}
  {330}},\ \bibinfo {pages} {1520} (\bibinfo {year} {2010})}\BibitemShut
  {NoStop}%
\bibitem [{\citenamefont {Safavi-Naeini}\ \emph {et~al.}(2011)\citenamefont
  {Safavi-Naeini}, \citenamefont {Alegre}, \citenamefont {Chan}, \citenamefont
  {Eichenfield}, \citenamefont {Winger}, \citenamefont {Lin}, \citenamefont
  {Hill}, \citenamefont {Chang},\ and\ \citenamefont
  {Painter}}]{safavi2011electromagnetically}%
  \BibitemOpen
  \bibfield  {author} {\bibinfo {author} {\bibfnamefont {A.~H.}\ \bibnamefont
  {Safavi-Naeini}}, \bibinfo {author} {\bibfnamefont {T.~M.}\ \bibnamefont
  {Alegre}}, \bibinfo {author} {\bibfnamefont {J.}~\bibnamefont {Chan}},
  \bibinfo {author} {\bibfnamefont {M.}~\bibnamefont {Eichenfield}}, \bibinfo
  {author} {\bibfnamefont {M.}~\bibnamefont {Winger}}, \bibinfo {author}
  {\bibfnamefont {Q.}~\bibnamefont {Lin}}, \bibinfo {author} {\bibfnamefont
  {J.~T.}\ \bibnamefont {Hill}}, \bibinfo {author} {\bibfnamefont {D.~E.}\
  \bibnamefont {Chang}}, \ and\ \bibinfo {author} {\bibfnamefont
  {O.}~\bibnamefont {Painter}},\ }\bibfield  {title} {\enquote {\bibinfo
  {title} {Electromagnetically induced transparency and slow light with
  optomechanics},}\ }\href@noop {} {\bibfield  {journal} {\bibinfo  {journal}
  {Nature}\ }\textbf {\bibinfo {volume} {472}},\ \bibinfo {pages} {69}
  (\bibinfo {year} {2011})}\BibitemShut {NoStop}%
\bibitem [{\citenamefont {Butsch}\ \emph {et~al.}(2014)\citenamefont {Butsch},
  \citenamefont {Koehler}, \citenamefont {Noskov},\ and\ \citenamefont
  {Russell}}]{butsch2014cw}%
  \BibitemOpen
  \bibfield  {author} {\bibinfo {author} {\bibfnamefont {A.}~\bibnamefont
  {Butsch}}, \bibinfo {author} {\bibfnamefont {J.}~\bibnamefont {Koehler}},
  \bibinfo {author} {\bibfnamefont {R.}~\bibnamefont {Noskov}}, \ and\ \bibinfo
  {author} {\bibfnamefont {P.~S.~J.}\ \bibnamefont {Russell}},\ }\bibfield
  {title} {\enquote {\bibinfo {title} {Cw-pumped single-pass frequency comb
  generation by resonant optomechanical nonlinearity in dual-nanoweb fiber},}\
  }\href@noop {} {\bibfield  {journal} {\bibinfo  {journal} {Optica}\ }\textbf
  {\bibinfo {volume} {1}},\ \bibinfo {pages} {158} (\bibinfo {year}
  {2014})}\BibitemShut {NoStop}%
\bibitem [{\citenamefont {Deotare}\ \emph {et~al.}(2012)\citenamefont
  {Deotare}, \citenamefont {Bulu}, \citenamefont {Frank}, \citenamefont {Quan},
  \citenamefont {Zhang}, \citenamefont {Ilic},\ and\ \citenamefont
  {Loncar}}]{deotare2012all}%
  \BibitemOpen
  \bibfield  {author} {\bibinfo {author} {\bibfnamefont {P.~B.}\ \bibnamefont
  {Deotare}}, \bibinfo {author} {\bibfnamefont {I.}~\bibnamefont {Bulu}},
  \bibinfo {author} {\bibfnamefont {I.~W.}\ \bibnamefont {Frank}}, \bibinfo
  {author} {\bibfnamefont {Q.}~\bibnamefont {Quan}}, \bibinfo {author}
  {\bibfnamefont {Y.}~\bibnamefont {Zhang}}, \bibinfo {author} {\bibfnamefont
  {R.}~\bibnamefont {Ilic}}, \ and\ \bibinfo {author} {\bibfnamefont
  {M.}~\bibnamefont {Loncar}},\ }\bibfield  {title} {\enquote {\bibinfo {title}
  {All optical reconfiguration of optomechanical filters},}\ }\href@noop {}
  {\bibfield  {journal} {\bibinfo  {journal} {Nature Communications}\ }\textbf
  {\bibinfo {volume} {3}},\ \bibinfo {pages} {846} (\bibinfo {year}
  {2012})}\BibitemShut {NoStop}%
\bibitem [{\citenamefont {Stannigel}\ \emph
  {et~al.}(2012{\natexlab{a}})\citenamefont {Stannigel}, \citenamefont {Komar},
  \citenamefont {Habraken}, \citenamefont {Bennett}, \citenamefont {Lukin},
  \citenamefont {Zoller},\ and\ \citenamefont
  {Rabl}}]{stannigel2012optomechanical}%
  \BibitemOpen
  \bibfield  {author} {\bibinfo {author} {\bibfnamefont {K.}~\bibnamefont
  {Stannigel}}, \bibinfo {author} {\bibfnamefont {P.}~\bibnamefont {Komar}},
  \bibinfo {author} {\bibfnamefont {S.}~\bibnamefont {Habraken}}, \bibinfo
  {author} {\bibfnamefont {S.}~\bibnamefont {Bennett}}, \bibinfo {author}
  {\bibfnamefont {M.~D.}\ \bibnamefont {Lukin}}, \bibinfo {author}
  {\bibfnamefont {P.}~\bibnamefont {Zoller}}, \ and\ \bibinfo {author}
  {\bibfnamefont {P.}~\bibnamefont {Rabl}},\ }\bibfield  {title} {\enquote
  {\bibinfo {title} {Optomechanical quantum information processing with photons
  and phonons},}\ }\href@noop {} {\bibfield  {journal} {\bibinfo  {journal}
  {Physical review letters}\ }\textbf {\bibinfo {volume} {109}},\ \bibinfo
  {pages} {013603} (\bibinfo {year} {2012}{\natexlab{a}})}\BibitemShut
  {NoStop}%
\bibitem [{\citenamefont {Vanner}\ \emph {et~al.}(2011)\citenamefont {Vanner},
  \citenamefont {Pikovski}, \citenamefont {Cole}, \citenamefont {Kim},
  \citenamefont {Brukner}, \citenamefont {Hammerer}, \citenamefont {Milburn},\
  and\ \citenamefont {Aspelmeyer}}]{vanner2011pulsed}%
  \BibitemOpen
  \bibfield  {author} {\bibinfo {author} {\bibfnamefont {M.~R.}\ \bibnamefont
  {Vanner}}, \bibinfo {author} {\bibfnamefont {I.}~\bibnamefont {Pikovski}},
  \bibinfo {author} {\bibfnamefont {G.~D.}\ \bibnamefont {Cole}}, \bibinfo
  {author} {\bibfnamefont {M.}~\bibnamefont {Kim}}, \bibinfo {author}
  {\bibfnamefont {{\v{C}}.}~\bibnamefont {Brukner}}, \bibinfo {author}
  {\bibfnamefont {K.}~\bibnamefont {Hammerer}}, \bibinfo {author}
  {\bibfnamefont {G.~J.}\ \bibnamefont {Milburn}}, \ and\ \bibinfo {author}
  {\bibfnamefont {M.}~\bibnamefont {Aspelmeyer}},\ }\bibfield  {title}
  {\enquote {\bibinfo {title} {Pulsed quantum optomechanics},}\ }\href@noop {}
  {\bibfield  {journal} {\bibinfo  {journal} {Proceedings of the National
  Academy of Sciences}\ }\textbf {\bibinfo {volume} {108}},\ \bibinfo {pages}
  {16182} (\bibinfo {year} {2011})}\BibitemShut {NoStop}%
\bibitem [{\citenamefont {Anetsberger}\ \emph {et~al.}(2010)\citenamefont
  {Anetsberger}, \citenamefont {Gavartin}, \citenamefont {Arcizet},
  \citenamefont {Unterreithmeier}, \citenamefont {Weig}, \citenamefont
  {Gorodetsky}, \citenamefont {Kotthaus},\ and\ \citenamefont
  {Kippenberg}}]{anetsberger2010measuring}%
  \BibitemOpen
  \bibfield  {author} {\bibinfo {author} {\bibfnamefont {G.}~\bibnamefont
  {Anetsberger}}, \bibinfo {author} {\bibfnamefont {E.}~\bibnamefont
  {Gavartin}}, \bibinfo {author} {\bibfnamefont {O.}~\bibnamefont {Arcizet}},
  \bibinfo {author} {\bibfnamefont {Q.~P.}\ \bibnamefont {Unterreithmeier}},
  \bibinfo {author} {\bibfnamefont {E.~M.}\ \bibnamefont {Weig}}, \bibinfo
  {author} {\bibfnamefont {M.~L.}\ \bibnamefont {Gorodetsky}}, \bibinfo
  {author} {\bibfnamefont {J.~P.}\ \bibnamefont {Kotthaus}}, \ and\ \bibinfo
  {author} {\bibfnamefont {T.~J.}\ \bibnamefont {Kippenberg}},\ }\bibfield
  {title} {\enquote {\bibinfo {title} {Measuring nanomechanical motion with an
  imprecision below the standard quantum limit},}\ }\href@noop {} {\bibfield
  {journal} {\bibinfo  {journal} {Physical Review A}\ }\textbf {\bibinfo
  {volume} {82}},\ \bibinfo {pages} {061804} (\bibinfo {year}
  {2010})}\BibitemShut {NoStop}%
\bibitem [{\citenamefont {Teufel}\ \emph {et~al.}(2011)\citenamefont {Teufel},
  \citenamefont {Donner}, \citenamefont {Li}, \citenamefont {Harlow},
  \citenamefont {Allman}, \citenamefont {Cicak}, \citenamefont {Sirois},
  \citenamefont {Whittaker}, \citenamefont {Lehnert},\ and\ \citenamefont
  {Simmonds}}]{teufel2011sideband}%
  \BibitemOpen
  \bibfield  {author} {\bibinfo {author} {\bibfnamefont {J.~D.}\ \bibnamefont
  {Teufel}}, \bibinfo {author} {\bibfnamefont {T.}~\bibnamefont {Donner}},
  \bibinfo {author} {\bibfnamefont {D.}~\bibnamefont {Li}}, \bibinfo {author}
  {\bibfnamefont {J.~W.}\ \bibnamefont {Harlow}}, \bibinfo {author}
  {\bibfnamefont {M.}~\bibnamefont {Allman}}, \bibinfo {author} {\bibfnamefont
  {K.}~\bibnamefont {Cicak}}, \bibinfo {author} {\bibfnamefont {A.~J.}\
  \bibnamefont {Sirois}}, \bibinfo {author} {\bibfnamefont {J.~D.}\
  \bibnamefont {Whittaker}}, \bibinfo {author} {\bibfnamefont {K.~W.}\
  \bibnamefont {Lehnert}}, \ and\ \bibinfo {author} {\bibfnamefont {R.~W.}\
  \bibnamefont {Simmonds}},\ }\bibfield  {title} {\enquote {\bibinfo {title}
  {Sideband cooling of micromechanical motion to the quantum ground state},}\
  }\href@noop {} {\bibfield  {journal} {\bibinfo  {journal} {Nature}\ }\textbf
  {\bibinfo {volume} {475}},\ \bibinfo {pages} {359} (\bibinfo {year}
  {2011})}\BibitemShut {NoStop}%
\bibitem [{\citenamefont {Meenehan}\ \emph {et~al.}(2015)\citenamefont
  {Meenehan}, \citenamefont {Cohen}, \citenamefont {MacCabe}, \citenamefont
  {Marsili}, \citenamefont {Shaw},\ and\ \citenamefont
  {Painter}}]{meenehan2015pulsed}%
  \BibitemOpen
  \bibfield  {author} {\bibinfo {author} {\bibfnamefont {S.~M.}\ \bibnamefont
  {Meenehan}}, \bibinfo {author} {\bibfnamefont {J.~D.}\ \bibnamefont {Cohen}},
  \bibinfo {author} {\bibfnamefont {G.~S.}\ \bibnamefont {MacCabe}}, \bibinfo
  {author} {\bibfnamefont {F.}~\bibnamefont {Marsili}}, \bibinfo {author}
  {\bibfnamefont {M.~D.}\ \bibnamefont {Shaw}}, \ and\ \bibinfo {author}
  {\bibfnamefont {O.}~\bibnamefont {Painter}},\ }\bibfield  {title} {\enquote
  {\bibinfo {title} {Pulsed excitation dynamics of an optomechanical crystal
  resonator near its quantum ground state of motion},}\ }\href@noop {}
  {\bibfield  {journal} {\bibinfo  {journal} {Physical Review X}\ }\textbf
  {\bibinfo {volume} {5}},\ \bibinfo {pages} {041002} (\bibinfo {year}
  {2015})}\BibitemShut {NoStop}%
\bibitem [{\citenamefont {Vitali}\ \emph {et~al.}(2007)\citenamefont {Vitali},
  \citenamefont {Gigan}, \citenamefont {Ferreira}, \citenamefont {B{\"o}hm},
  \citenamefont {Tombesi}, \citenamefont {Guerreiro}, \citenamefont {Vedral},
  \citenamefont {Zeilinger},\ and\ \citenamefont
  {Aspelmeyer}}]{vitali2007optomechanical}%
  \BibitemOpen
  \bibfield  {author} {\bibinfo {author} {\bibfnamefont {D.}~\bibnamefont
  {Vitali}}, \bibinfo {author} {\bibfnamefont {S.}~\bibnamefont {Gigan}},
  \bibinfo {author} {\bibfnamefont {A.}~\bibnamefont {Ferreira}}, \bibinfo
  {author} {\bibfnamefont {H.}~\bibnamefont {B{\"o}hm}}, \bibinfo {author}
  {\bibfnamefont {P.}~\bibnamefont {Tombesi}}, \bibinfo {author} {\bibfnamefont
  {A.}~\bibnamefont {Guerreiro}}, \bibinfo {author} {\bibfnamefont
  {V.}~\bibnamefont {Vedral}}, \bibinfo {author} {\bibfnamefont
  {A.}~\bibnamefont {Zeilinger}}, \ and\ \bibinfo {author} {\bibfnamefont
  {M.}~\bibnamefont {Aspelmeyer}},\ }\bibfield  {title} {\enquote {\bibinfo
  {title} {Optomechanical entanglement between a movable mirror and a cavity
  field},}\ }\href@noop {} {\bibfield  {journal} {\bibinfo  {journal} {Physical
  review letters}\ }\textbf {\bibinfo {volume} {98}},\ \bibinfo {pages}
  {030405} (\bibinfo {year} {2007})}\BibitemShut {NoStop}%
\bibitem [{\citenamefont {Yang}\ \emph {et~al.}(2019)\citenamefont {Yang},
  \citenamefont {Yin},\ and\ \citenamefont {Xiao}}]{yang2019optomechanically}%
  \BibitemOpen
  \bibfield  {author} {\bibinfo {author} {\bibfnamefont {X.}~\bibnamefont
  {Yang}}, \bibinfo {author} {\bibfnamefont {Z.}~\bibnamefont {Yin}}, \ and\
  \bibinfo {author} {\bibfnamefont {M.}~\bibnamefont {Xiao}},\ }\bibfield
  {title} {\enquote {\bibinfo {title} {Optomechanically induced
  entanglement},}\ }\href@noop {} {\bibfield  {journal} {\bibinfo  {journal}
  {Physical Review A}\ }\textbf {\bibinfo {volume} {99}},\ \bibinfo {pages}
  {013811} (\bibinfo {year} {2019})}\BibitemShut {NoStop}%
\bibitem [{\citenamefont {L{\"u}}\ \emph {et~al.}(2015)\citenamefont {L{\"u}},
  \citenamefont {Liao}, \citenamefont {Tian},\ and\ \citenamefont
  {Nori}}]{lu2015steady}%
  \BibitemOpen
  \bibfield  {author} {\bibinfo {author} {\bibfnamefont {X.-Y.}\ \bibnamefont
  {L{\"u}}}, \bibinfo {author} {\bibfnamefont {J.-Q.}\ \bibnamefont {Liao}},
  \bibinfo {author} {\bibfnamefont {L.}~\bibnamefont {Tian}}, \ and\ \bibinfo
  {author} {\bibfnamefont {F.}~\bibnamefont {Nori}},\ }\bibfield  {title}
  {\enquote {\bibinfo {title} {Steady-state mechanical squeezing in an
  optomechanical system via duffing nonlinearity},}\ }\href@noop {} {\bibfield
  {journal} {\bibinfo  {journal} {Physical Review A}\ }\textbf {\bibinfo
  {volume} {91}},\ \bibinfo {pages} {013834} (\bibinfo {year}
  {2015})}\BibitemShut {NoStop}%
\bibitem [{\citenamefont {Branford}\ \emph {et~al.}(2018)\citenamefont
  {Branford}, \citenamefont {Miao},\ and\ \citenamefont
  {Datta}}]{branford2018fundamental}%
  \BibitemOpen
  \bibfield  {author} {\bibinfo {author} {\bibfnamefont {D.}~\bibnamefont
  {Branford}}, \bibinfo {author} {\bibfnamefont {H.}~\bibnamefont {Miao}}, \
  and\ \bibinfo {author} {\bibfnamefont {A.}~\bibnamefont {Datta}},\ }\bibfield
   {title} {\enquote {\bibinfo {title} {Fundamental quantum limits of
  multicarrier optomechanical sensors},}\ }\href@noop {} {\bibfield  {journal}
  {\bibinfo  {journal} {Physical review letters}\ }\textbf {\bibinfo {volume}
  {121}},\ \bibinfo {pages} {110505} (\bibinfo {year} {2018})}\BibitemShut
  {NoStop}%
\bibitem [{\citenamefont {Arcizet}\ \emph {et~al.}(2006)\citenamefont
  {Arcizet}, \citenamefont {Cohadon}, \citenamefont {Briant}, \citenamefont
  {Pinard}, \citenamefont {Heidmann}, \citenamefont {Mackowski}, \citenamefont
  {Michel}, \citenamefont {Pinard}, \citenamefont {Fran{\c{c}}ais},\ and\
  \citenamefont {Rousseau}}]{arcizet2006high}%
  \BibitemOpen
  \bibfield  {author} {\bibinfo {author} {\bibfnamefont {O.}~\bibnamefont
  {Arcizet}}, \bibinfo {author} {\bibfnamefont {P.-F.}\ \bibnamefont
  {Cohadon}}, \bibinfo {author} {\bibfnamefont {T.}~\bibnamefont {Briant}},
  \bibinfo {author} {\bibfnamefont {M.}~\bibnamefont {Pinard}}, \bibinfo
  {author} {\bibfnamefont {A.}~\bibnamefont {Heidmann}}, \bibinfo {author}
  {\bibfnamefont {J.-M.}\ \bibnamefont {Mackowski}}, \bibinfo {author}
  {\bibfnamefont {C.}~\bibnamefont {Michel}}, \bibinfo {author} {\bibfnamefont
  {L.}~\bibnamefont {Pinard}}, \bibinfo {author} {\bibfnamefont
  {O.}~\bibnamefont {Fran{\c{c}}ais}}, \ and\ \bibinfo {author} {\bibfnamefont
  {L.}~\bibnamefont {Rousseau}},\ }\bibfield  {title} {\enquote {\bibinfo
  {title} {High-sensitivity optical monitoring of a micromechanical resonator
  with a quantum-limited optomechanical sensor},}\ }\href@noop {} {\bibfield
  {journal} {\bibinfo  {journal} {Physical review letters}\ }\textbf {\bibinfo
  {volume} {97}},\ \bibinfo {pages} {133601} (\bibinfo {year}
  {2006})}\BibitemShut {NoStop}%
\bibitem [{\citenamefont {Lee}\ \emph {et~al.}(2015)\citenamefont {Lee},
  \citenamefont {Underwood}, \citenamefont {Mason}, \citenamefont {Shkarin},
  \citenamefont {Hoch},\ and\ \citenamefont {Harris}}]{lee2015multimode}%
  \BibitemOpen
  \bibfield  {author} {\bibinfo {author} {\bibfnamefont {D.}~\bibnamefont
  {Lee}}, \bibinfo {author} {\bibfnamefont {M.}~\bibnamefont {Underwood}},
  \bibinfo {author} {\bibfnamefont {D.}~\bibnamefont {Mason}}, \bibinfo
  {author} {\bibfnamefont {A.}~\bibnamefont {Shkarin}}, \bibinfo {author}
  {\bibfnamefont {S.}~\bibnamefont {Hoch}}, \ and\ \bibinfo {author}
  {\bibfnamefont {J.}~\bibnamefont {Harris}},\ }\bibfield  {title} {\enquote
  {\bibinfo {title} {Multimode optomechanical dynamics in a cavity with avoided
  crossings},}\ }\href@noop {} {\bibfield  {journal} {\bibinfo  {journal}
  {Nature communications}\ }\textbf {\bibinfo {volume} {6}},\ \bibinfo {pages}
  {6232} (\bibinfo {year} {2015})}\BibitemShut {NoStop}%
\bibitem [{\citenamefont {Fan}\ \emph {et~al.}(2015)\citenamefont {Fan},
  \citenamefont {Fong}, \citenamefont {Poot},\ and\ \citenamefont
  {Tang}}]{fan2015cascaded}%
  \BibitemOpen
  \bibfield  {author} {\bibinfo {author} {\bibfnamefont {L.}~\bibnamefont
  {Fan}}, \bibinfo {author} {\bibfnamefont {K.~Y.}\ \bibnamefont {Fong}},
  \bibinfo {author} {\bibfnamefont {M.}~\bibnamefont {Poot}}, \ and\ \bibinfo
  {author} {\bibfnamefont {H.~X.}\ \bibnamefont {Tang}},\ }\bibfield  {title}
  {\enquote {\bibinfo {title} {Cascaded optical transparency in
  multimode-cavity optomechanical systems},}\ }\href@noop {} {\bibfield
  {journal} {\bibinfo  {journal} {Nature communications}\ }\textbf {\bibinfo
  {volume} {6}},\ \bibinfo {pages} {5850} (\bibinfo {year} {2015})}\BibitemShut
  {NoStop}%
\bibitem [{\citenamefont {Duggan}\ \emph {et~al.}(2019)\citenamefont {Duggan},
  \citenamefont {del Pino}, \citenamefont {Verhagen},\ and\ \citenamefont
  {Al{\`u}}}]{duggan2019optomechanically}%
  \BibitemOpen
  \bibfield  {author} {\bibinfo {author} {\bibfnamefont {R.}~\bibnamefont
  {Duggan}}, \bibinfo {author} {\bibfnamefont {J.}~\bibnamefont {del Pino}},
  \bibinfo {author} {\bibfnamefont {E.}~\bibnamefont {Verhagen}}, \ and\
  \bibinfo {author} {\bibfnamefont {A.}~\bibnamefont {Al{\`u}}},\ }\bibfield
  {title} {\enquote {\bibinfo {title} {Optomechanically induced birefringence
  and optomechanically induced faraday effect},}\ }\href@noop {} {\bibfield
  {journal} {\bibinfo  {journal} {Physical Review Letters}\ }\textbf {\bibinfo
  {volume} {123}},\ \bibinfo {pages} {023602} (\bibinfo {year}
  {2019})}\BibitemShut {NoStop}%
\bibitem [{\citenamefont {Wei}\ \emph {et~al.}(2019)\citenamefont {Wei},
  \citenamefont {Sheng}, \citenamefont {Yang}, \citenamefont {Wu},\ and\
  \citenamefont {Wu}}]{wei2019controllable}%
  \BibitemOpen
  \bibfield  {author} {\bibinfo {author} {\bibfnamefont {X.}~\bibnamefont
  {Wei}}, \bibinfo {author} {\bibfnamefont {J.}~\bibnamefont {Sheng}}, \bibinfo
  {author} {\bibfnamefont {C.}~\bibnamefont {Yang}}, \bibinfo {author}
  {\bibfnamefont {Y.}~\bibnamefont {Wu}}, \ and\ \bibinfo {author}
  {\bibfnamefont {H.}~\bibnamefont {Wu}},\ }\bibfield  {title} {\enquote
  {\bibinfo {title} {Controllable two-membrane-in-the-middle cavity
  optomechanical system},}\ }\href@noop {} {\bibfield  {journal} {\bibinfo
  {journal} {Physical Review A}\ }\textbf {\bibinfo {volume} {99}},\ \bibinfo
  {pages} {023851} (\bibinfo {year} {2019})}\BibitemShut {NoStop}%
\bibitem [{\citenamefont {Shen}\ \emph {et~al.}(2016)\citenamefont {Shen},
  \citenamefont {Zhang}, \citenamefont {Chen}, \citenamefont {Zou},
  \citenamefont {Xiao}, \citenamefont {Zou}, \citenamefont {Sun}, \citenamefont
  {Guo},\ and\ \citenamefont {Dong}}]{shen2016experimental}%
  \BibitemOpen
  \bibfield  {author} {\bibinfo {author} {\bibfnamefont {Z.}~\bibnamefont
  {Shen}}, \bibinfo {author} {\bibfnamefont {Y.-L.}\ \bibnamefont {Zhang}},
  \bibinfo {author} {\bibfnamefont {Y.}~\bibnamefont {Chen}}, \bibinfo {author}
  {\bibfnamefont {C.-L.}\ \bibnamefont {Zou}}, \bibinfo {author} {\bibfnamefont
  {Y.-F.}\ \bibnamefont {Xiao}}, \bibinfo {author} {\bibfnamefont {X.-B.}\
  \bibnamefont {Zou}}, \bibinfo {author} {\bibfnamefont {F.-W.}\ \bibnamefont
  {Sun}}, \bibinfo {author} {\bibfnamefont {G.-C.}\ \bibnamefont {Guo}}, \ and\
  \bibinfo {author} {\bibfnamefont {C.-H.}\ \bibnamefont {Dong}},\ }\bibfield
  {title} {\enquote {\bibinfo {title} {Experimental realization of
  optomechanically induced non-reciprocity},}\ }\href@noop {} {\bibfield
  {journal} {\bibinfo  {journal} {Nature Photonics}\ }\textbf {\bibinfo
  {volume} {10}},\ \bibinfo {pages} {657} (\bibinfo {year} {2016})}\BibitemShut
  {NoStop}%
\bibitem [{\citenamefont {Ruesink}\ \emph {et~al.}(2018)\citenamefont
  {Ruesink}, \citenamefont {Mathew}, \citenamefont {Miri}, \citenamefont
  {Al{\`u}},\ and\ \citenamefont {Verhagen}}]{ruesink2018optical}%
  \BibitemOpen
  \bibfield  {author} {\bibinfo {author} {\bibfnamefont {F.}~\bibnamefont
  {Ruesink}}, \bibinfo {author} {\bibfnamefont {J.~P.}\ \bibnamefont {Mathew}},
  \bibinfo {author} {\bibfnamefont {M.-A.}\ \bibnamefont {Miri}}, \bibinfo
  {author} {\bibfnamefont {A.}~\bibnamefont {Al{\`u}}}, \ and\ \bibinfo
  {author} {\bibfnamefont {E.}~\bibnamefont {Verhagen}},\ }\bibfield  {title}
  {\enquote {\bibinfo {title} {Optical circulation in a multimode
  optomechanical resonator},}\ }\href@noop {} {\bibfield  {journal} {\bibinfo
  {journal} {Nature communications}\ }\textbf {\bibinfo {volume} {9}},\
  \bibinfo {pages} {1798} (\bibinfo {year} {2018})}\BibitemShut {NoStop}%
\bibitem [{\citenamefont {Shen}\ \emph {et~al.}(2018)\citenamefont {Shen},
  \citenamefont {Zhang}, \citenamefont {Chen}, \citenamefont {Sun},
  \citenamefont {Zou}, \citenamefont {Guo}, \citenamefont {Zou},\ and\
  \citenamefont {Dong}}]{shen2018reconfigurable}%
  \BibitemOpen
  \bibfield  {author} {\bibinfo {author} {\bibfnamefont {Z.}~\bibnamefont
  {Shen}}, \bibinfo {author} {\bibfnamefont {Y.-L.}\ \bibnamefont {Zhang}},
  \bibinfo {author} {\bibfnamefont {Y.}~\bibnamefont {Chen}}, \bibinfo {author}
  {\bibfnamefont {F.-W.}\ \bibnamefont {Sun}}, \bibinfo {author} {\bibfnamefont
  {X.-B.}\ \bibnamefont {Zou}}, \bibinfo {author} {\bibfnamefont {G.-C.}\
  \bibnamefont {Guo}}, \bibinfo {author} {\bibfnamefont {C.-L.}\ \bibnamefont
  {Zou}}, \ and\ \bibinfo {author} {\bibfnamefont {C.-H.}\ \bibnamefont
  {Dong}},\ }\bibfield  {title} {\enquote {\bibinfo {title} {Reconfigurable
  optomechanical circulator and directional amplifier},}\ }\href@noop {}
  {\bibfield  {journal} {\bibinfo  {journal} {Nature communications}\ }\textbf
  {\bibinfo {volume} {9}},\ \bibinfo {pages} {1797} (\bibinfo {year}
  {2018})}\BibitemShut {NoStop}%
\bibitem [{\citenamefont {Dong}\ \emph {et~al.}(2012)\citenamefont {Dong},
  \citenamefont {Fiore}, \citenamefont {Kuzyk},\ and\ \citenamefont
  {Wang}}]{dong2012optomechanical}%
  \BibitemOpen
  \bibfield  {author} {\bibinfo {author} {\bibfnamefont {C.}~\bibnamefont
  {Dong}}, \bibinfo {author} {\bibfnamefont {V.}~\bibnamefont {Fiore}},
  \bibinfo {author} {\bibfnamefont {M.~C.}\ \bibnamefont {Kuzyk}}, \ and\
  \bibinfo {author} {\bibfnamefont {H.}~\bibnamefont {Wang}},\ }\bibfield
  {title} {\enquote {\bibinfo {title} {Optomechanical dark mode},}\ }\href@noop
  {} {\bibfield  {journal} {\bibinfo  {journal} {Science}\ }\textbf {\bibinfo
  {volume} {338}},\ \bibinfo {pages} {1609} (\bibinfo {year}
  {2012})}\BibitemShut {NoStop}%
\bibitem [{\citenamefont {Hill}\ \emph {et~al.}(2012)\citenamefont {Hill},
  \citenamefont {Safavi-Naeini}, \citenamefont {Chan},\ and\ \citenamefont
  {Painter}}]{hill2012coherent}%
  \BibitemOpen
  \bibfield  {author} {\bibinfo {author} {\bibfnamefont {J.~T.}\ \bibnamefont
  {Hill}}, \bibinfo {author} {\bibfnamefont {A.~H.}\ \bibnamefont
  {Safavi-Naeini}}, \bibinfo {author} {\bibfnamefont {J.}~\bibnamefont {Chan}},
  \ and\ \bibinfo {author} {\bibfnamefont {O.}~\bibnamefont {Painter}},\
  }\bibfield  {title} {\enquote {\bibinfo {title} {Coherent optical wavelength
  conversion via cavity optomechanics},}\ }\href@noop {} {\bibfield  {journal}
  {\bibinfo  {journal} {Nature communications}\ }\textbf {\bibinfo {volume}
  {3}},\ \bibinfo {pages} {1196} (\bibinfo {year} {2012})}\BibitemShut
  {NoStop}%
\bibitem [{\citenamefont {Verstraete}\ \emph {et~al.}(2009)\citenamefont
  {Verstraete}, \citenamefont {Wolf},\ and\ \citenamefont
  {Cirac}}]{verstraete2009quantum}%
  \BibitemOpen
  \bibfield  {author} {\bibinfo {author} {\bibfnamefont {F.}~\bibnamefont
  {Verstraete}}, \bibinfo {author} {\bibfnamefont {M.~M.}\ \bibnamefont
  {Wolf}}, \ and\ \bibinfo {author} {\bibfnamefont {J.~I.}\ \bibnamefont
  {Cirac}},\ }\bibfield  {title} {\enquote {\bibinfo {title} {Quantum
  computation and quantum-state engineering driven by dissipation},}\
  }\href@noop {} {\bibfield  {journal} {\bibinfo  {journal} {Nature physics}\
  }\textbf {\bibinfo {volume} {5}},\ \bibinfo {pages} {633} (\bibinfo {year}
  {2009})}\BibitemShut {NoStop}%
\bibitem [{\citenamefont {Poyatos}\ \emph {et~al.}(1996)\citenamefont
  {Poyatos}, \citenamefont {Cirac},\ and\ \citenamefont
  {Zoller}}]{poyatos1996quantum}%
  \BibitemOpen
  \bibfield  {author} {\bibinfo {author} {\bibfnamefont {J.}~\bibnamefont
  {Poyatos}}, \bibinfo {author} {\bibfnamefont {J.~I.}\ \bibnamefont {Cirac}},
  \ and\ \bibinfo {author} {\bibfnamefont {P.}~\bibnamefont {Zoller}},\
  }\bibfield  {title} {\enquote {\bibinfo {title} {Quantum reservoir
  engineering with laser cooled trapped ions},}\ }\href@noop {} {\bibfield
  {journal} {\bibinfo  {journal} {Physical review letters}\ }\textbf {\bibinfo
  {volume} {77}},\ \bibinfo {pages} {4728} (\bibinfo {year}
  {1996})}\BibitemShut {NoStop}%
\bibitem [{\citenamefont {Stannigel}\ \emph
  {et~al.}(2012{\natexlab{b}})\citenamefont {Stannigel}, \citenamefont {Rabl},\
  and\ \citenamefont {Zoller}}]{stannigel2012driven}%
  \BibitemOpen
  \bibfield  {author} {\bibinfo {author} {\bibfnamefont {K.}~\bibnamefont
  {Stannigel}}, \bibinfo {author} {\bibfnamefont {P.}~\bibnamefont {Rabl}}, \
  and\ \bibinfo {author} {\bibfnamefont {P.}~\bibnamefont {Zoller}},\
  }\bibfield  {title} {\enquote {\bibinfo {title} {Driven-dissipative
  preparation of entangled states in cascaded quantum-optical networks},}\
  }\href@noop {} {\bibfield  {journal} {\bibinfo  {journal} {New Journal of
  Physics}\ }\textbf {\bibinfo {volume} {14}},\ \bibinfo {pages} {063014}
  (\bibinfo {year} {2012}{\natexlab{b}})}\BibitemShut {NoStop}%
\bibitem [{\citenamefont {Sarlette}\ \emph {et~al.}(2011)\citenamefont
  {Sarlette}, \citenamefont {Raimond}, \citenamefont {Brune},\ and\
  \citenamefont {Rouchon}}]{sarlette2011stabilization}%
  \BibitemOpen
  \bibfield  {author} {\bibinfo {author} {\bibfnamefont {A.}~\bibnamefont
  {Sarlette}}, \bibinfo {author} {\bibfnamefont {J.-M.}\ \bibnamefont
  {Raimond}}, \bibinfo {author} {\bibfnamefont {M.}~\bibnamefont {Brune}}, \
  and\ \bibinfo {author} {\bibfnamefont {P.}~\bibnamefont {Rouchon}},\
  }\bibfield  {title} {\enquote {\bibinfo {title} {Stabilization of
  nonclassical states of the radiation field in a cavity by reservoir
  engineering},}\ }\href@noop {} {\bibfield  {journal} {\bibinfo  {journal}
  {Physical review letters}\ }\textbf {\bibinfo {volume} {107}},\ \bibinfo
  {pages} {010402} (\bibinfo {year} {2011})}\BibitemShut {NoStop}%
\bibitem [{\citenamefont {Leghtas}\ \emph {et~al.}(2013)\citenamefont
  {Leghtas}, \citenamefont {Vool}, \citenamefont {Shankar}, \citenamefont
  {Hatridge}, \citenamefont {Girvin}, \citenamefont {Devoret},\ and\
  \citenamefont {Mirrahimi}}]{leghtas2013stabilizing}%
  \BibitemOpen
  \bibfield  {author} {\bibinfo {author} {\bibfnamefont {Z.}~\bibnamefont
  {Leghtas}}, \bibinfo {author} {\bibfnamefont {U.}~\bibnamefont {Vool}},
  \bibinfo {author} {\bibfnamefont {S.}~\bibnamefont {Shankar}}, \bibinfo
  {author} {\bibfnamefont {M.}~\bibnamefont {Hatridge}}, \bibinfo {author}
  {\bibfnamefont {S.~M.}\ \bibnamefont {Girvin}}, \bibinfo {author}
  {\bibfnamefont {M.~H.}\ \bibnamefont {Devoret}}, \ and\ \bibinfo {author}
  {\bibfnamefont {M.}~\bibnamefont {Mirrahimi}},\ }\bibfield  {title} {\enquote
  {\bibinfo {title} {Stabilizing a bell state of two superconducting qubits by
  dissipation engineering},}\ }\href@noop {} {\bibfield  {journal} {\bibinfo
  {journal} {Physical Review A}\ }\textbf {\bibinfo {volume} {88}},\ \bibinfo
  {pages} {023849} (\bibinfo {year} {2013})}\BibitemShut {NoStop}%
\bibitem [{\citenamefont {Ma}\ \emph {et~al.}(2019)\citenamefont {Ma},
  \citenamefont {Li}, \citenamefont {Liu}, \citenamefont {Xie},\ and\
  \citenamefont {Li}}]{ma2019stabilizing}%
  \BibitemOpen
  \bibfield  {author} {\bibinfo {author} {\bibfnamefont {S.-l.}\ \bibnamefont
  {Ma}}, \bibinfo {author} {\bibfnamefont {X.-k.}\ \bibnamefont {Li}}, \bibinfo
  {author} {\bibfnamefont {X.-y.}\ \bibnamefont {Liu}}, \bibinfo {author}
  {\bibfnamefont {J.-k.}\ \bibnamefont {Xie}}, \ and\ \bibinfo {author}
  {\bibfnamefont {F.-l.}\ \bibnamefont {Li}},\ }\bibfield  {title} {\enquote
  {\bibinfo {title} {Stabilizing bell states of two separated superconducting
  qubits via quantum reservoir engineering},}\ }\href@noop {} {\bibfield
  {journal} {\bibinfo  {journal} {Physical Review A}\ }\textbf {\bibinfo
  {volume} {99}},\ \bibinfo {pages} {042336} (\bibinfo {year}
  {2019})}\BibitemShut {NoStop}%
\bibitem [{\citenamefont {Barreiro}\ \emph {et~al.}(2011)\citenamefont
  {Barreiro}, \citenamefont {M{\"u}ller}, \citenamefont {Schindler},
  \citenamefont {Nigg}, \citenamefont {Monz}, \citenamefont {Chwalla},
  \citenamefont {Hennrich}, \citenamefont {Roos}, \citenamefont {Zoller},\ and\
  \citenamefont {Blatt}}]{barreiro2011open}%
  \BibitemOpen
  \bibfield  {author} {\bibinfo {author} {\bibfnamefont {J.~T.}\ \bibnamefont
  {Barreiro}}, \bibinfo {author} {\bibfnamefont {M.}~\bibnamefont
  {M{\"u}ller}}, \bibinfo {author} {\bibfnamefont {P.}~\bibnamefont
  {Schindler}}, \bibinfo {author} {\bibfnamefont {D.}~\bibnamefont {Nigg}},
  \bibinfo {author} {\bibfnamefont {T.}~\bibnamefont {Monz}}, \bibinfo {author}
  {\bibfnamefont {M.}~\bibnamefont {Chwalla}}, \bibinfo {author} {\bibfnamefont
  {M.}~\bibnamefont {Hennrich}}, \bibinfo {author} {\bibfnamefont {C.~F.}\
  \bibnamefont {Roos}}, \bibinfo {author} {\bibfnamefont {P.}~\bibnamefont
  {Zoller}}, \ and\ \bibinfo {author} {\bibfnamefont {R.}~\bibnamefont
  {Blatt}},\ }\bibfield  {title} {\enquote {\bibinfo {title} {An open-system
  quantum simulator with trapped ions},}\ }\href@noop {} {\bibfield  {journal}
  {\bibinfo  {journal} {Nature}\ }\textbf {\bibinfo {volume} {470}},\ \bibinfo
  {pages} {486} (\bibinfo {year} {2011})}\BibitemShut {NoStop}%
\bibitem [{\citenamefont {Schindler}\ \emph {et~al.}(2013)\citenamefont
  {Schindler}, \citenamefont {M{\"u}ller}, \citenamefont {Nigg}, \citenamefont
  {Barreiro}, \citenamefont {Martinez}, \citenamefont {Hennrich}, \citenamefont
  {Monz}, \citenamefont {Diehl}, \citenamefont {Zoller},\ and\ \citenamefont
  {Blatt}}]{schindler2013quantum}%
  \BibitemOpen
  \bibfield  {author} {\bibinfo {author} {\bibfnamefont {P.}~\bibnamefont
  {Schindler}}, \bibinfo {author} {\bibfnamefont {M.}~\bibnamefont
  {M{\"u}ller}}, \bibinfo {author} {\bibfnamefont {D.}~\bibnamefont {Nigg}},
  \bibinfo {author} {\bibfnamefont {J.~T.}\ \bibnamefont {Barreiro}}, \bibinfo
  {author} {\bibfnamefont {E.~A.}\ \bibnamefont {Martinez}}, \bibinfo {author}
  {\bibfnamefont {M.}~\bibnamefont {Hennrich}}, \bibinfo {author}
  {\bibfnamefont {T.}~\bibnamefont {Monz}}, \bibinfo {author} {\bibfnamefont
  {S.}~\bibnamefont {Diehl}}, \bibinfo {author} {\bibfnamefont
  {P.}~\bibnamefont {Zoller}}, \ and\ \bibinfo {author} {\bibfnamefont
  {R.}~\bibnamefont {Blatt}},\ }\bibfield  {title} {\enquote {\bibinfo {title}
  {Quantum simulation of dynamical maps with trapped ions},}\ }\href@noop {}
  {\bibfield  {journal} {\bibinfo  {journal} {Nature Physics}\ }\textbf
  {\bibinfo {volume} {9}},\ \bibinfo {pages} {361} (\bibinfo {year}
  {2013})}\BibitemShut {NoStop}%
\bibitem [{\citenamefont {Li}\ \emph {et~al.}(2012)\citenamefont {Li},
  \citenamefont {Gao},\ and\ \citenamefont {Li}}]{li2012engineering}%
  \BibitemOpen
  \bibfield  {author} {\bibinfo {author} {\bibfnamefont {P.-B.}\ \bibnamefont
  {Li}}, \bibinfo {author} {\bibfnamefont {S.-Y.}\ \bibnamefont {Gao}}, \ and\
  \bibinfo {author} {\bibfnamefont {F.-L.}\ \bibnamefont {Li}},\ }\bibfield
  {title} {\enquote {\bibinfo {title} {Engineering two-mode entangled states
  between two superconducting resonators by dissipation},}\ }\href@noop {}
  {\bibfield  {journal} {\bibinfo  {journal} {Physical Review A}\ }\textbf
  {\bibinfo {volume} {86}},\ \bibinfo {pages} {012318} (\bibinfo {year}
  {2012})}\BibitemShut {NoStop}%
\bibitem [{\citenamefont {Li}\ \emph {et~al.}(2018)\citenamefont {Li},
  \citenamefont {Shao}, \citenamefont {Wu}, \citenamefont {Yi},\ and\
  \citenamefont {Zheng}}]{li2018engineering}%
  \BibitemOpen
  \bibfield  {author} {\bibinfo {author} {\bibfnamefont {D.-X.}\ \bibnamefont
  {Li}}, \bibinfo {author} {\bibfnamefont {X.-Q.}\ \bibnamefont {Shao}},
  \bibinfo {author} {\bibfnamefont {J.-H.}\ \bibnamefont {Wu}}, \bibinfo
  {author} {\bibfnamefont {X.}~\bibnamefont {Yi}}, \ and\ \bibinfo {author}
  {\bibfnamefont {T.-Y.}\ \bibnamefont {Zheng}},\ }\bibfield  {title} {\enquote
  {\bibinfo {title} {Engineering steady knill-laflamme-milburn state of rydberg
  atoms by dissipation},}\ }\href@noop {} {\bibfield  {journal} {\bibinfo
  {journal} {Optics express}\ }\textbf {\bibinfo {volume} {26}},\ \bibinfo
  {pages} {2292} (\bibinfo {year} {2018})}\BibitemShut {NoStop}%
\bibitem [{\citenamefont {Massel}\ \emph {et~al.}(2012)\citenamefont {Massel},
  \citenamefont {Cho}, \citenamefont {Pirkkalainen}, \citenamefont {Hakonen},
  \citenamefont {Heikkil{\"a}},\ and\ \citenamefont
  {Sillanp{\"a}{\"a}}}]{massel2012multimode}%
  \BibitemOpen
  \bibfield  {author} {\bibinfo {author} {\bibfnamefont {F.}~\bibnamefont
  {Massel}}, \bibinfo {author} {\bibfnamefont {S.~U.}\ \bibnamefont {Cho}},
  \bibinfo {author} {\bibfnamefont {J.-M.}\ \bibnamefont {Pirkkalainen}},
  \bibinfo {author} {\bibfnamefont {P.~J.}\ \bibnamefont {Hakonen}}, \bibinfo
  {author} {\bibfnamefont {T.~T.}\ \bibnamefont {Heikkil{\"a}}}, \ and\
  \bibinfo {author} {\bibfnamefont {M.~A.}\ \bibnamefont {Sillanp{\"a}{\"a}}},\
  }\bibfield  {title} {\enquote {\bibinfo {title} {Multimode circuit
  optomechanics near the quantum limit},}\ }\href@noop {} {\bibfield  {journal}
  {\bibinfo  {journal} {Nature communications}\ }\textbf {\bibinfo {volume}
  {3}},\ \bibinfo {pages} {987} (\bibinfo {year} {2012})}\BibitemShut {NoStop}%
\bibitem [{\citenamefont {Deng}\ \emph {et~al.}(2016)\citenamefont {Deng},
  \citenamefont {Yan}, \citenamefont {Wang},\ and\ \citenamefont
  {Wu}}]{deng2016optimizing}%
  \BibitemOpen
  \bibfield  {author} {\bibinfo {author} {\bibfnamefont {Z.~J.}\ \bibnamefont
  {Deng}}, \bibinfo {author} {\bibfnamefont {X.-B.}\ \bibnamefont {Yan}},
  \bibinfo {author} {\bibfnamefont {Y.-D.}\ \bibnamefont {Wang}}, \ and\
  \bibinfo {author} {\bibfnamefont {C.-W.}\ \bibnamefont {Wu}},\ }\bibfield
  {title} {\enquote {\bibinfo {title} {Optimizing the output-photon
  entanglement in multimode optomechanical systems},}\ }\href@noop {}
  {\bibfield  {journal} {\bibinfo  {journal} {Physical Review A}\ }\textbf
  {\bibinfo {volume} {93}},\ \bibinfo {pages} {033842} (\bibinfo {year}
  {2016})}\BibitemShut {NoStop}%
\bibitem [{\citenamefont {Ockeloen-Korppi}\ \emph {et~al.}(2019)\citenamefont
  {Ockeloen-Korppi}, \citenamefont {Gely}, \citenamefont {Damsk{\"a}gg},
  \citenamefont {Jenkins}, \citenamefont {Steele},\ and\ \citenamefont
  {Sillanp{\"a}{\"a}}}]{ockeloen2019sideband}%
  \BibitemOpen
  \bibfield  {author} {\bibinfo {author} {\bibfnamefont {C.}~\bibnamefont
  {Ockeloen-Korppi}}, \bibinfo {author} {\bibfnamefont {M.}~\bibnamefont
  {Gely}}, \bibinfo {author} {\bibfnamefont {E.}~\bibnamefont {Damsk{\"a}gg}},
  \bibinfo {author} {\bibfnamefont {M.}~\bibnamefont {Jenkins}}, \bibinfo
  {author} {\bibfnamefont {G.}~\bibnamefont {Steele}}, \ and\ \bibinfo {author}
  {\bibfnamefont {M.}~\bibnamefont {Sillanp{\"a}{\"a}}},\ }\bibfield  {title}
  {\enquote {\bibinfo {title} {Sideband cooling of nearly degenerate
  micromechanical oscillators in a multimode optomechanical system},}\
  }\href@noop {} {\bibfield  {journal} {\bibinfo  {journal} {Physical Review
  A}\ }\textbf {\bibinfo {volume} {99}},\ \bibinfo {pages} {023826} (\bibinfo
  {year} {2019})}\BibitemShut {NoStop}%
\bibitem [{\citenamefont {Zhang}\ \emph {et~al.}(2017)\citenamefont {Zhang},
  \citenamefont {Dong}, \citenamefont {Zou}, \citenamefont {Zou}, \citenamefont
  {Wang},\ and\ \citenamefont {Guo}}]{zhang2017optomechanical}%
  \BibitemOpen
  \bibfield  {author} {\bibinfo {author} {\bibfnamefont {Y.-L.}\ \bibnamefont
  {Zhang}}, \bibinfo {author} {\bibfnamefont {C.-H.}\ \bibnamefont {Dong}},
  \bibinfo {author} {\bibfnamefont {C.-L.}\ \bibnamefont {Zou}}, \bibinfo
  {author} {\bibfnamefont {X.-B.}\ \bibnamefont {Zou}}, \bibinfo {author}
  {\bibfnamefont {Y.-D.}\ \bibnamefont {Wang}}, \ and\ \bibinfo {author}
  {\bibfnamefont {G.-C.}\ \bibnamefont {Guo}},\ }\bibfield  {title} {\enquote
  {\bibinfo {title} {Optomechanical devices based on traveling-wave
  microresonators},}\ }\href@noop {} {\bibfield  {journal} {\bibinfo  {journal}
  {Physical Review A}\ }\textbf {\bibinfo {volume} {95}},\ \bibinfo {pages}
  {043815} (\bibinfo {year} {2017})}\BibitemShut {NoStop}%
\bibitem [{\citenamefont {Nielsen}\ \emph {et~al.}(2017)\citenamefont
  {Nielsen}, \citenamefont {Tsaturyan}, \citenamefont {M{\o}ller},
  \citenamefont {Polzik},\ and\ \citenamefont
  {Schliesser}}]{nielsen2017multimode}%
  \BibitemOpen
  \bibfield  {author} {\bibinfo {author} {\bibfnamefont {W.~H.~P.}\
  \bibnamefont {Nielsen}}, \bibinfo {author} {\bibfnamefont {Y.}~\bibnamefont
  {Tsaturyan}}, \bibinfo {author} {\bibfnamefont {C.~B.}\ \bibnamefont
  {M{\o}ller}}, \bibinfo {author} {\bibfnamefont {E.~S.}\ \bibnamefont
  {Polzik}}, \ and\ \bibinfo {author} {\bibfnamefont {A.}~\bibnamefont
  {Schliesser}},\ }\bibfield  {title} {\enquote {\bibinfo {title} {Multimode
  optomechanical system in the quantum regime},}\ }\href@noop {} {\bibfield
  {journal} {\bibinfo  {journal} {Proceedings of the National Academy of
  Sciences}\ }\textbf {\bibinfo {volume} {114}},\ \bibinfo {pages} {62}
  (\bibinfo {year} {2017})}\BibitemShut {NoStop}%
\bibitem [{\citenamefont {Meystre}(2013)}]{meystre2013short}%
  \BibitemOpen
  \bibfield  {author} {\bibinfo {author} {\bibfnamefont {P.}~\bibnamefont
  {Meystre}},\ }\bibfield  {title} {\enquote {\bibinfo {title} {A short walk
  through quantum optomechanics},}\ }\href@noop {} {\bibfield  {journal}
  {\bibinfo  {journal} {Annalen der Physik}\ }\textbf {\bibinfo {volume}
  {525}},\ \bibinfo {pages} {215} (\bibinfo {year} {2013})}\BibitemShut
  {NoStop}%
\bibitem [{\citenamefont {Lee}\ and\ \citenamefont
  {Seok}(2018)}]{lee2018quantum}%
  \BibitemOpen
  \bibfield  {author} {\bibinfo {author} {\bibfnamefont {J.~H.}\ \bibnamefont
  {Lee}}\ and\ \bibinfo {author} {\bibfnamefont {H.}~\bibnamefont {Seok}},\
  }\bibfield  {title} {\enquote {\bibinfo {title} {Quantum reservoir
  engineering through quadratic optomechanical interaction in the reversed
  dissipation regime},}\ }\href@noop {} {\bibfield  {journal} {\bibinfo
  {journal} {Physical Review A}\ }\textbf {\bibinfo {volume} {97}},\ \bibinfo
  {pages} {013805} (\bibinfo {year} {2018})}\BibitemShut {NoStop}%
\bibitem [{\citenamefont {Chen}\ \emph {et~al.}(2017)\citenamefont {Chen},
  \citenamefont {Liao},\ and\ \citenamefont {Lin}}]{chen2017dissipative}%
  \BibitemOpen
  \bibfield  {author} {\bibinfo {author} {\bibfnamefont {R.-X.}\ \bibnamefont
  {Chen}}, \bibinfo {author} {\bibfnamefont {C.-G.}\ \bibnamefont {Liao}}, \
  and\ \bibinfo {author} {\bibfnamefont {X.-M.}\ \bibnamefont {Lin}},\
  }\bibfield  {title} {\enquote {\bibinfo {title} {Dissipative generation of
  significant amount of mechanical entanglement in a coupled optomechanical
  system},}\ }\href@noop {} {\bibfield  {journal} {\bibinfo  {journal}
  {Scientific Reports}\ }\textbf {\bibinfo {volume} {7}},\ \bibinfo {pages}
  {14497} (\bibinfo {year} {2017})}\BibitemShut {NoStop}%
\bibitem [{\citenamefont {Hu}\ \emph {et~al.}(2019)\citenamefont {Hu},
  \citenamefont {Zhao}, \citenamefont {Wang}, \citenamefont {Zhang},
  \citenamefont {Zou}, \citenamefont {Dong}, \citenamefont {Tang},
  \citenamefont {Guo},\ and\ \citenamefont {Zou}}]{hu2019cavity}%
  \BibitemOpen
  \bibfield  {author} {\bibinfo {author} {\bibfnamefont {X.-X.}\ \bibnamefont
  {Hu}}, \bibinfo {author} {\bibfnamefont {C.-L.}\ \bibnamefont {Zhao}},
  \bibinfo {author} {\bibfnamefont {Z.-B.}\ \bibnamefont {Wang}}, \bibinfo
  {author} {\bibfnamefont {Y.-L.}\ \bibnamefont {Zhang}}, \bibinfo {author}
  {\bibfnamefont {X.-B.}\ \bibnamefont {Zou}}, \bibinfo {author} {\bibfnamefont
  {C.-H.}\ \bibnamefont {Dong}}, \bibinfo {author} {\bibfnamefont {H.~X.}\
  \bibnamefont {Tang}}, \bibinfo {author} {\bibfnamefont {G.-C.}\ \bibnamefont
  {Guo}}, \ and\ \bibinfo {author} {\bibfnamefont {C.-L.}\ \bibnamefont
  {Zou}},\ }\bibfield  {title} {\enquote {\bibinfo {title} {Cavity-enhanced
  optical controlling based on three-wave mixing in cavity-atom ensemble
  system},}\ }\href@noop {} {\bibfield  {journal} {\bibinfo  {journal} {Optics
  express}\ }\textbf {\bibinfo {volume} {27}},\ \bibinfo {pages} {6660}
  (\bibinfo {year} {2019})}\BibitemShut {NoStop}%
\bibitem [{\citenamefont {Devoret}\ and\ \citenamefont
  {Schoelkopf}(2013)}]{devoret2013superconducting}%
  \BibitemOpen
  \bibfield  {author} {\bibinfo {author} {\bibfnamefont {M.~H.}\ \bibnamefont
  {Devoret}}\ and\ \bibinfo {author} {\bibfnamefont {R.~J.}\ \bibnamefont
  {Schoelkopf}},\ }\bibfield  {title} {\enquote {\bibinfo {title}
  {Superconducting circuits for quantum information: an outlook},}\ }\href@noop
  {} {\bibfield  {journal} {\bibinfo  {journal} {Science}\ }\textbf {\bibinfo
  {volume} {339}},\ \bibinfo {pages} {1169} (\bibinfo {year}
  {2013})}\BibitemShut {NoStop}%
\bibitem [{\citenamefont {Wallquist}\ \emph {et~al.}(2009)\citenamefont
  {Wallquist}, \citenamefont {Hammerer}, \citenamefont {Rabl}, \citenamefont
  {Lukin},\ and\ \citenamefont {Zoller}}]{wallquist2009hybrid}%
  \BibitemOpen
  \bibfield  {author} {\bibinfo {author} {\bibfnamefont {M.}~\bibnamefont
  {Wallquist}}, \bibinfo {author} {\bibfnamefont {K.}~\bibnamefont {Hammerer}},
  \bibinfo {author} {\bibfnamefont {P.}~\bibnamefont {Rabl}}, \bibinfo {author}
  {\bibfnamefont {M.}~\bibnamefont {Lukin}}, \ and\ \bibinfo {author}
  {\bibfnamefont {P.}~\bibnamefont {Zoller}},\ }\bibfield  {title} {\enquote
  {\bibinfo {title} {Hybrid quantum devices and quantum engineering},}\
  }\href@noop {} {\bibfield  {journal} {\bibinfo  {journal} {Physica Scripta}\
  }\textbf {\bibinfo {volume} {2009}},\ \bibinfo {pages} {014001} (\bibinfo
  {year} {2009})}\BibitemShut {NoStop}%
\bibitem [{\citenamefont {Wang}\ and\ \citenamefont
  {Clerk}(2013)}]{wang2013reservoir}%
  \BibitemOpen
  \bibfield  {author} {\bibinfo {author} {\bibfnamefont {Y.-D.}\ \bibnamefont
  {Wang}}\ and\ \bibinfo {author} {\bibfnamefont {A.~A.}\ \bibnamefont
  {Clerk}},\ }\bibfield  {title} {\enquote {\bibinfo {title}
  {Reservoir-engineered entanglement in optomechanical systems},}\ }\href@noop
  {} {\bibfield  {journal} {\bibinfo  {journal} {Physical review letters}\
  }\textbf {\bibinfo {volume} {110}},\ \bibinfo {pages} {253601} (\bibinfo
  {year} {2013})}\BibitemShut {NoStop}%
\bibitem [{\citenamefont {Tian}(2013)}]{tian2013robust}%
  \BibitemOpen
  \bibfield  {author} {\bibinfo {author} {\bibfnamefont {L.}~\bibnamefont
  {Tian}},\ }\bibfield  {title} {\enquote {\bibinfo {title} {Robust photon
  entanglement via quantum interference in optomechanical interfaces},}\
  }\href@noop {} {\bibfield  {journal} {\bibinfo  {journal} {Physical review
  letters}\ }\textbf {\bibinfo {volume} {110}},\ \bibinfo {pages} {233602}
  (\bibinfo {year} {2013})}\BibitemShut {NoStop}%
\bibitem [{\citenamefont {Hartmann}\ and\ \citenamefont
  {Plenio}(2008)}]{hartmann2008steady}%
  \BibitemOpen
  \bibfield  {author} {\bibinfo {author} {\bibfnamefont {M.~J.}\ \bibnamefont
  {Hartmann}}\ and\ \bibinfo {author} {\bibfnamefont {M.~B.}\ \bibnamefont
  {Plenio}},\ }\bibfield  {title} {\enquote {\bibinfo {title} {Steady state
  entanglement in the mechanical vibrations of two dielectric membranes},}\
  }\href@noop {} {\bibfield  {journal} {\bibinfo  {journal} {Physical Review
  Letters}\ }\textbf {\bibinfo {volume} {101}},\ \bibinfo {pages} {200503}
  (\bibinfo {year} {2008})}\BibitemShut {NoStop}%
\bibitem [{\citenamefont {Zhang}\ \emph {et~al.}(2016)\citenamefont {Zhang},
  \citenamefont {Zhu}, \citenamefont {Zou},\ and\ \citenamefont
  {Tang}}]{zhang2016optomagnonic}%
  \BibitemOpen
  \bibfield  {author} {\bibinfo {author} {\bibfnamefont {X.}~\bibnamefont
  {Zhang}}, \bibinfo {author} {\bibfnamefont {N.}~\bibnamefont {Zhu}}, \bibinfo
  {author} {\bibfnamefont {C.-L.}\ \bibnamefont {Zou}}, \ and\ \bibinfo
  {author} {\bibfnamefont {H.~X.}\ \bibnamefont {Tang}},\ }\bibfield  {title}
  {\enquote {\bibinfo {title} {Optomagnonic whispering gallery
  microresonators},}\ }\href@noop {} {\bibfield  {journal} {\bibinfo  {journal}
  {Physical review letters}\ }\textbf {\bibinfo {volume} {117}},\ \bibinfo
  {pages} {123605} (\bibinfo {year} {2016})}\BibitemShut {NoStop}%
\bibitem [{\citenamefont {Pezz{\`e}}\ \emph {et~al.}(2018)\citenamefont
  {Pezz{\`e}}, \citenamefont {Smerzi}, \citenamefont {Oberthaler},
  \citenamefont {Schmied},\ and\ \citenamefont {Treutlein}}]{pezze2018quantum}%
  \BibitemOpen
  \bibfield  {author} {\bibinfo {author} {\bibfnamefont {L.}~\bibnamefont
  {Pezz{\`e}}}, \bibinfo {author} {\bibfnamefont {A.}~\bibnamefont {Smerzi}},
  \bibinfo {author} {\bibfnamefont {M.~K.}\ \bibnamefont {Oberthaler}},
  \bibinfo {author} {\bibfnamefont {R.}~\bibnamefont {Schmied}}, \ and\
  \bibinfo {author} {\bibfnamefont {P.}~\bibnamefont {Treutlein}},\ }\bibfield
  {title} {\enquote {\bibinfo {title} {Quantum metrology with nonclassical
  states of atomic ensembles},}\ }\href@noop {} {\bibfield  {journal} {\bibinfo
   {journal} {Reviews of Modern Physics}\ }\textbf {\bibinfo {volume} {90}},\
  \bibinfo {pages} {035005} (\bibinfo {year} {2018})}\BibitemShut {NoStop}%
\bibitem [{\citenamefont {Viasnoff-Schwoob}\ \emph {et~al.}(2005)\citenamefont
  {Viasnoff-Schwoob}, \citenamefont {Weisbuch}, \citenamefont {Benisty},
  \citenamefont {Cuisin}, \citenamefont {Derouin}, \citenamefont {Drisse},
  \citenamefont {Duan}, \citenamefont {Legouezigou}, \citenamefont
  {Legouezigou}, \citenamefont {Pommereau} \emph
  {et~al.}}]{viasnoff2005compact}%
  \BibitemOpen
  \bibfield  {author} {\bibinfo {author} {\bibfnamefont {E.}~\bibnamefont
  {Viasnoff-Schwoob}}, \bibinfo {author} {\bibfnamefont {C.}~\bibnamefont
  {Weisbuch}}, \bibinfo {author} {\bibfnamefont {H.}~\bibnamefont {Benisty}},
  \bibinfo {author} {\bibfnamefont {C.}~\bibnamefont {Cuisin}}, \bibinfo
  {author} {\bibfnamefont {E.}~\bibnamefont {Derouin}}, \bibinfo {author}
  {\bibfnamefont {O.}~\bibnamefont {Drisse}}, \bibinfo {author} {\bibfnamefont
  {G.~H.}\ \bibnamefont {Duan}}, \bibinfo {author} {\bibfnamefont
  {L.}~\bibnamefont {Legouezigou}}, \bibinfo {author} {\bibfnamefont
  {O.}~\bibnamefont {Legouezigou}}, \bibinfo {author} {\bibfnamefont
  {F.}~\bibnamefont {Pommereau}},  \emph {et~al.},\ }\bibfield  {title}
  {\enquote {\bibinfo {title} {Compact wavelength monitoring by lateral
  outcoupling in wedged photonic crystal multimode waveguides},}\ }\href@noop
  {} {\bibfield  {journal} {\bibinfo  {journal} {Applied Physics Letters}\
  }\textbf {\bibinfo {volume} {86}},\ \bibinfo {pages} {101107} (\bibinfo
  {year} {2005})}\BibitemShut {NoStop}%
\bibitem [{\citenamefont {Dong}\ \emph {et~al.}(2015)\citenamefont {Dong},
  \citenamefont {Shen}, \citenamefont {Zou}, \citenamefont {Zhang},
  \citenamefont {Fu},\ and\ \citenamefont {Guo}}]{dong2015brillouin}%
  \BibitemOpen
  \bibfield  {author} {\bibinfo {author} {\bibfnamefont {C.-H.}\ \bibnamefont
  {Dong}}, \bibinfo {author} {\bibfnamefont {Z.}~\bibnamefont {Shen}}, \bibinfo
  {author} {\bibfnamefont {C.-L.}\ \bibnamefont {Zou}}, \bibinfo {author}
  {\bibfnamefont {Y.-L.}\ \bibnamefont {Zhang}}, \bibinfo {author}
  {\bibfnamefont {W.}~\bibnamefont {Fu}}, \ and\ \bibinfo {author}
  {\bibfnamefont {G.-C.}\ \bibnamefont {Guo}},\ }\bibfield  {title} {\enquote
  {\bibinfo {title} {Brillouin-scattering-induced transparency and
  non-reciprocal light storage},}\ }\href@noop {} {\bibfield  {journal}
  {\bibinfo  {journal} {Nature communications}\ }\textbf {\bibinfo {volume}
  {6}},\ \bibinfo {pages} {6193} (\bibinfo {year} {2015})}\BibitemShut
  {NoStop}%
\bibitem [{\citenamefont {Grudinin}\ \emph {et~al.}(2010)\citenamefont
  {Grudinin}, \citenamefont {Lee}, \citenamefont {Painter},\ and\ \citenamefont
  {Vahala}}]{grudinin2010phonon}%
  \BibitemOpen
  \bibfield  {author} {\bibinfo {author} {\bibfnamefont {I.~S.}\ \bibnamefont
  {Grudinin}}, \bibinfo {author} {\bibfnamefont {H.}~\bibnamefont {Lee}},
  \bibinfo {author} {\bibfnamefont {O.}~\bibnamefont {Painter}}, \ and\
  \bibinfo {author} {\bibfnamefont {K.~J.}\ \bibnamefont {Vahala}},\ }\bibfield
   {title} {\enquote {\bibinfo {title} {Phonon laser action in a tunable
  two-level system},}\ }\href@noop {} {\bibfield  {journal} {\bibinfo
  {journal} {Physical review letters}\ }\textbf {\bibinfo {volume} {104}},\
  \bibinfo {pages} {083901} (\bibinfo {year} {2010})}\BibitemShut {NoStop}%
\bibitem [{\citenamefont {Weedbrook}\ \emph {et~al.}(2012)\citenamefont
  {Weedbrook}, \citenamefont {Pirandola}, \citenamefont
  {Garc{\'\i}a-Patr{\'o}n}, \citenamefont {Cerf}, \citenamefont {Ralph},
  \citenamefont {Shapiro},\ and\ \citenamefont
  {Lloyd}}]{weedbrook2012gaussian}%
  \BibitemOpen
  \bibfield  {author} {\bibinfo {author} {\bibfnamefont {C.}~\bibnamefont
  {Weedbrook}}, \bibinfo {author} {\bibfnamefont {S.}~\bibnamefont
  {Pirandola}}, \bibinfo {author} {\bibfnamefont {R.}~\bibnamefont
  {Garc{\'\i}a-Patr{\'o}n}}, \bibinfo {author} {\bibfnamefont {N.~J.}\
  \bibnamefont {Cerf}}, \bibinfo {author} {\bibfnamefont {T.~C.}\ \bibnamefont
  {Ralph}}, \bibinfo {author} {\bibfnamefont {J.~H.}\ \bibnamefont {Shapiro}},
  \ and\ \bibinfo {author} {\bibfnamefont {S.}~\bibnamefont {Lloyd}},\
  }\bibfield  {title} {\enquote {\bibinfo {title} {Gaussian quantum
  information},}\ }\href@noop {} {\bibfield  {journal} {\bibinfo  {journal}
  {Reviews of Modern Physics}\ }\textbf {\bibinfo {volume} {84}},\ \bibinfo
  {pages} {621} (\bibinfo {year} {2012})}\BibitemShut {NoStop}%
\bibitem [{\citenamefont {Plenio}(2005)}]{plenio2005logarithmic}%
  \BibitemOpen
  \bibfield  {author} {\bibinfo {author} {\bibfnamefont {M.~B.}\ \bibnamefont
  {Plenio}},\ }\bibfield  {title} {\enquote {\bibinfo {title} {Logarithmic
  negativity: a full entanglement monotone that is not convex},}\ }\href@noop
  {} {\bibfield  {journal} {\bibinfo  {journal} {Physical review letters}\
  }\textbf {\bibinfo {volume} {95}},\ \bibinfo {pages} {090503} (\bibinfo
  {year} {2005})}\BibitemShut {NoStop}%
\bibitem [{\citenamefont {DeJesus}\ and\ \citenamefont
  {Kaufman}(1987)}]{dejesus1987routh}%
  \BibitemOpen
  \bibfield  {author} {\bibinfo {author} {\bibfnamefont {E.~X.}\ \bibnamefont
  {DeJesus}}\ and\ \bibinfo {author} {\bibfnamefont {C.}~\bibnamefont
  {Kaufman}},\ }\bibfield  {title} {\enquote {\bibinfo {title} {Routh-hurwitz
  criterion in the examination of eigenvalues of a system of nonlinear ordinary
  differential equations},}\ }\href@noop {} {\bibfield  {journal} {\bibinfo
  {journal} {Physical Review A}\ }\textbf {\bibinfo {volume} {35}},\ \bibinfo
  {pages} {5288} (\bibinfo {year} {1987})}\BibitemShut {NoStop}%
\end{thebibliography}%

\end{document}